\documentstyle[preprint,tighten,aps,floats,colordvi,epsfig]{revtex}

\begin{document}
\draft
\vskip 2cm

\title{Large Wilson loops with overlap and clover fermions: Two-loop evaluation of the b-quark mass shift and the quark-antiquark potential}

\author{A. Athenodorou and H. Panagopoulos}
\address{Department of Physics, University of Cyprus, P.O. Box 20537,
Lefkosia CY-1678, Cyprus \\
{\it email: }{\tt ph00aa1@ucy.ac.cy, haris@ucy.ac.cy}}
\vskip 3mm

\date{\today}

\maketitle

\begin{abstract}

We compute, to two loops in pertubation theory, the fermionic contribution to
 rectangular $R\times T$ Wilson loops, for different values of $R$ and $T$.

We use the overlap fermionic action. We also employ the clover action,
 for comparison with existing results in the literature.
 In the limit $R, T \rightarrow \infty$ our results lead to the shift in the b-quark mass.
We also evaluate the perturbative static potential as $T \rightarrow \infty$.

 
\end{abstract}

\eject

\section{Introduction}
\label{Introduction}

In recent years there has been a great progress toward the implementation of chirality preserving regularizations of gauge theories coupled to fermions on the lattice. The two most widely studied approaches are the overlap fermions and domain wall fermions. Numerical simulations and perturbative calculations, using these actions, are already giving promising results.

 In our paper, we compute the perturbative value of \WildStrawberry{{\bf large Wilson loops}} up to two loops, using the \RedViolet{clover} (SW) and \RedViolet{overlap} fermions. Using the perturbative values of Wilson loops 
of infinite length, we evaluate the shift of the \WildStrawberry{{\bf b-quark mass}}. The perturbative values of Wilson loops, of infinite time extent, lead us to the evaluation of the \WildStrawberry{{\bf static potential}}. For the case of clover fermions, we also compare our results with established results.

The calculation of Wilson loops in lattice perturbation theory is useful in a number of ways: It leads to the prediction of a strong coupling constant 
\Blue{$a_{\overline{MS}}\left( m_  
Z \right)$} from low energy hadronic phenomenology by means of non-perturbative lattice simulations~\cite{AXEGHAK,CTDKH1,CTDKH2,ASEA,SBEA,CDEA}. It is employed in the context of mean field improvement programmes 
of the lattice action and operators~\cite{GP,GPLPBM}. In the limit of infinite time separation, 
$T \rightarrow \infty$, Wilson loops give access to the perturbative 
quark-antiquark potential. Furthermore in the limit of large distances, 
$R \rightarrow \infty$, the self energy of static sources can be 
obtained from the potential, enabling the calculation of 
 \Blue{$\overline{m_b}\left(\overline{m_b}\right)$} from the non-perturbative 
simulations of heavy-light mesons in the static limit~\cite{GMCTS}.

\section{Calculation of Wilson loops}
\label{Calculation of Wilson loops}

The Wilson loop $W$, around a closed curve $C$, is the expectation 
value of the path ordered product of gauge links:

\begin{equation}
W\left( C \right)=\left\langle tr \left[ \prod_{x_i \in C} U_{x_i,\mu_i} \right] \right\rangle,
\end{equation} 

\noindent
where $\mu_i \in {\pm1,\cdots,\pm4}$, denotes the direction 
indicated by $x_{i+1}-x_i$ and $U_{x,-\mu}=U_{x-\widehat{\mu},\mu}$. $W(R,T)$, 
denotes a rectangular Wilson loop where the closed curve contains two opposite 
lines with an extent of $T$ lattice units pointing in the time direction, separated by a spatial distance $aR$.

The smallest Wilson loop is the plaquette $\Box$, which is the expectation 
value of the Wilson gauge action:

%%Clover-Wilson Action%%
\begin{eqnarray}
 S_W &=& \beta \sum_\Box  \left[1 - {1\over N} \hbox{Re Tr} 
\left( \Box \right)\right].
\end{eqnarray}

There are two Feynman diagrams involving fermions, contributing to $W\left(R,T\right)$ at two loops, as shown in Fig. 1:

\begin{center}
\psfig{width=8.5truecm,file=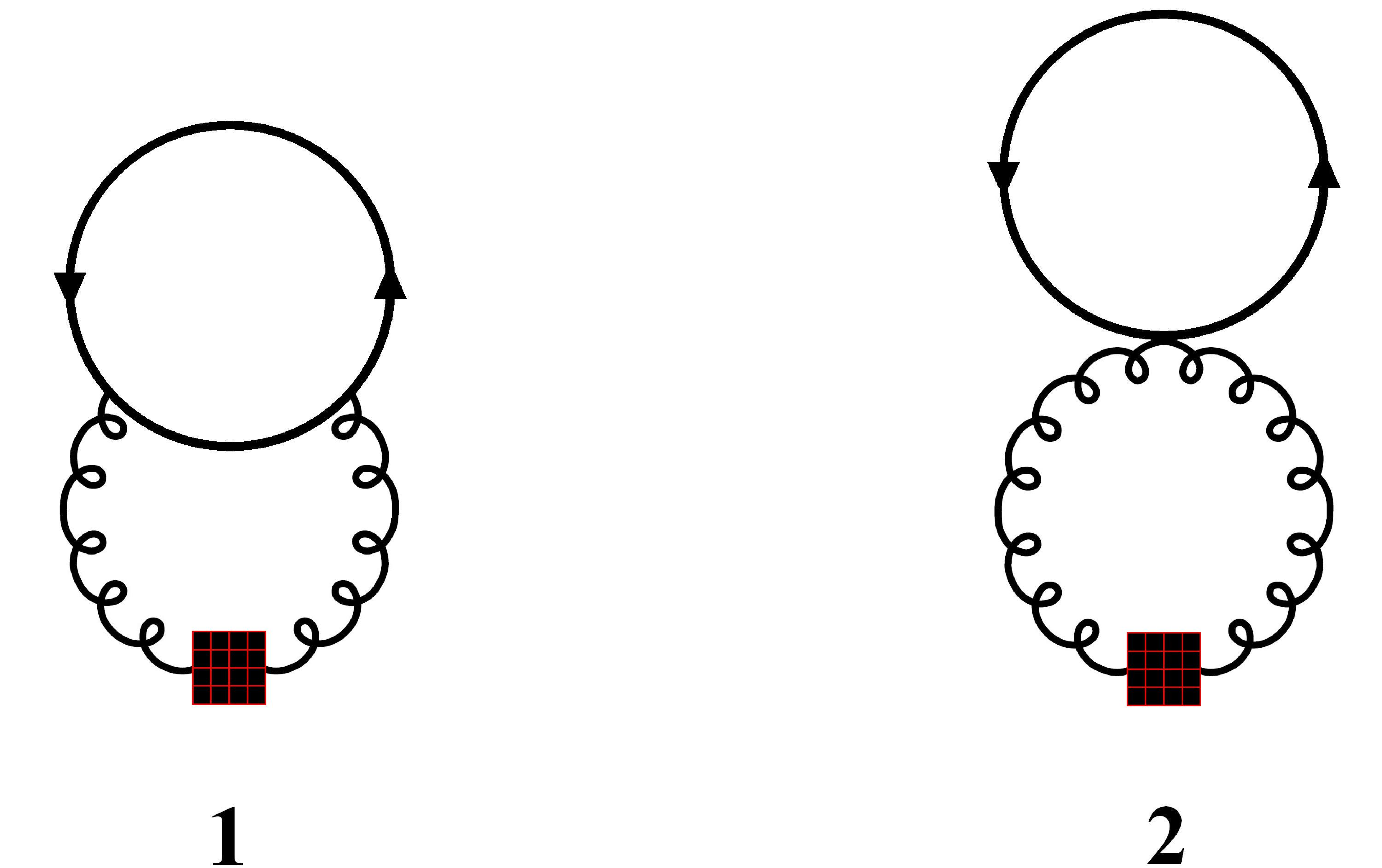}

{\footnotesize Fig.1: Two-loop fermionic diagrams contributing to $W\left(R,T\right)$.} 
\end{center}

\noindent
The grid-like square vertex stands for two - point vertex of $W\left(R,T\right)$, whose 
mathematical expression is:

\begin{eqnarray}
\sum_x &tr& \left(\prod_{x_i \in C} U_{x_i,\mu_i} \right)\rightarrow -{g^2\over24}\sum_{\mu, \nu}
\int {d^4kd^4k'\over\left(2 \pi\right)^4} A^a_{\mu}\left(k\right)A^b_{\nu}\left(k'\right) \nonumber\\
&\times& \delta^{ab}\delta\left(k+k'\right) \Bigg[2\delta_{\mu,\nu}{\rm S} \left(k_{\mu}, R\right) \sum_{\rho}\sin^2\left(k_{\rho}aT/2\right)\\
&+& 2\delta_{\mu,\nu} {\rm S} \left(k_{\mu}, T\right)\sum_{\rho}\sin^2\left(k_{\rho}aR/2\right)\nonumber\\ 
&-& 4 {\rm S} \left(k_{\mu}, R\right) {\rm S} \left(k_{\nu}, T\right)\sin\left(k_{\mu}a/2\right)\sin\left(k_{\nu}a/2\right)\Bigg],\nonumber 
\end{eqnarray}

\noindent
where: S$\left(k_{\mu}, R\right)\equiv{\displaystyle{\sin^2\left(Rk_{\mu}a/2\right)}/\displaystyle{\sin^2\left(k_{\mu}a/2\right)}}$. The involved algebra of the lattice perturbation theory was carried out using our computer package in Mathematica. The value of each diagram is computed numerically for a 
sequence of finite lattice sizes. Their values have been summed, and then extrapolated to infinite lattice size.

\section{Calculation with Clover Fermions}
\label{Calculation with Clover Fermions}

The action is, in standard notation:
\begin{eqnarray}
  S &=& S_W + S_f ,
\end{eqnarray}
where:
\begin{eqnarray}
 S_f &=& \sum_{x,f} 
         \Bigg[ \left(m + {4 r} \right)
         \overline\psi^f_x \psi^f_x -  \\ 
      & & {1 \over 2} \sum_\mu
         \left( \overline\psi^f_{x+\hat\mu} \left( r +
         \gamma_\mu \right) U^\dagger_\mu(x) \psi^f_x +
         \overline\psi^f_x \left(r - \gamma_\mu \right)
         U_\mu(x) \psi^f_{x+\hat\mu} \right) \Bigg]  \nonumber\\
&+& {i\over 4}\,\Blue{c_{\rm SW}}\,\sum_f\sum_{\mu,\,\nu} \overline\psi^f_x
\sigma_{\mu\nu} {\hat F}^{\mu\nu}_x \psi^f_x ,\nonumber
\end{eqnarray}

\noindent
$$
{\rm and:}\quad {\hat F}^{\mu\nu} \equiv {1\over8}\,
(Q_{\mu\nu} - Q_{\nu\mu}), \quad 
Q_{\mu\nu} = U_{\mu,\nu} {+} U_{\nu,{-}\mu} {+} U_{{-}\mu,{-}\nu} {+} 
U_{{-}\nu,\mu}
$$

Here $U_{\mu,\,\nu}(x)$ is the usual product of link variables
$U_{\mu}(x)$ along a plaquette in the $\mu$-$\nu$
directions, originating at $x$;
$f$ is a flavor index; $m$ is the
bare fermionic mass; $\sigma_{\mu\nu} =(i/2) [\gamma_\mu,\,\gamma_\nu]$; powers of $a$
may be directly reinserted by dimensional counting.
The clover coefficient \Blue{$c_{\rm SW}$ is a free parameter} in the present
work; it is normally 
tuned in a way as to minimize ${\cal O}(a)$ effects.

The perturbative expansion of the Wilson loop is given 
by the expression:

\begin{equation}
W\left(C\right)\, /\, N=1-W_{LO}\ g^2-\left(W_{NLO}-{\left(N^2-1\right)\over N}N_f\Red{X}\right)g^4,
\label{Wc}
\end{equation}

\noindent
where: $W_{LO}$ and $W_{NLO}$ are the pure gauge contributions with the Wilson gauge action, which can be found in Ref.~\cite{GBPB,DRLS}, and:

\begin{equation}
\Red{X}=\Red{X_W}+\Red{X^a_{SW}}\Blue{c_{SW}}+\Red{X^b_{SW}}\Blue{c^2_{SW}}.
\label{Xi}
\end{equation}

We have computed the values of $\Red{X_W}$, $\Red{X^a_{SW}}$, and $\Red{X^b_{SW}}$, and we list them in Tables \ref{CloverSw0}, \ref{CloverSw1}, \ref{CloverSw2}. We compare our results for $\Red{X_{W}}$, $\Red{X^a_{SW}}$, and $\Red{X^b_{SW}}$  
with those of Ref.~\cite{DRLS}\footnote{In order to compare with Ref.~\cite{DRLS}, we deduce their values of $X_{W}$ from the data presented there and we estimate the errors stemming from that data.} (only $X_{W}$), 
and Ref.~\cite{GBPB} (for $m=0$). 
The comparison can be found in Tables \ref{Comp1}, \ref{Comp2}, \ref{Comp3}.

Our results for $\Red{X_W}$, $\Red{X^a_{SW}}$, and $\Red{X^b_{SW}}$ as a 
function of mass for square loops, are shown in Figs. 2-4.

\begin{center}
\psfig{width=15truecm,file=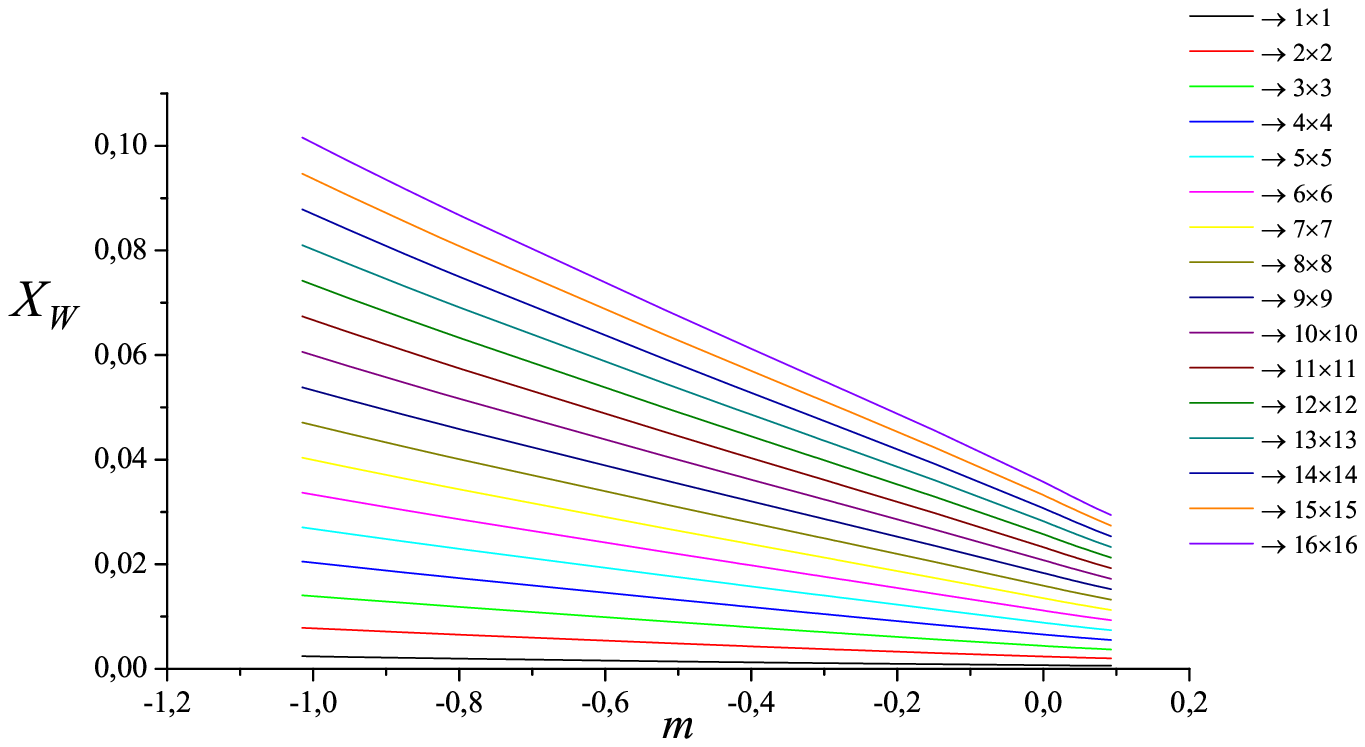}

{\footnotesize Fig.2: $X_W$ for loops $L\times L$. Top line: $L=16$, bottom line: $L=1$.}
\end{center}

\begin{center}
\psfig{width=15truecm,file=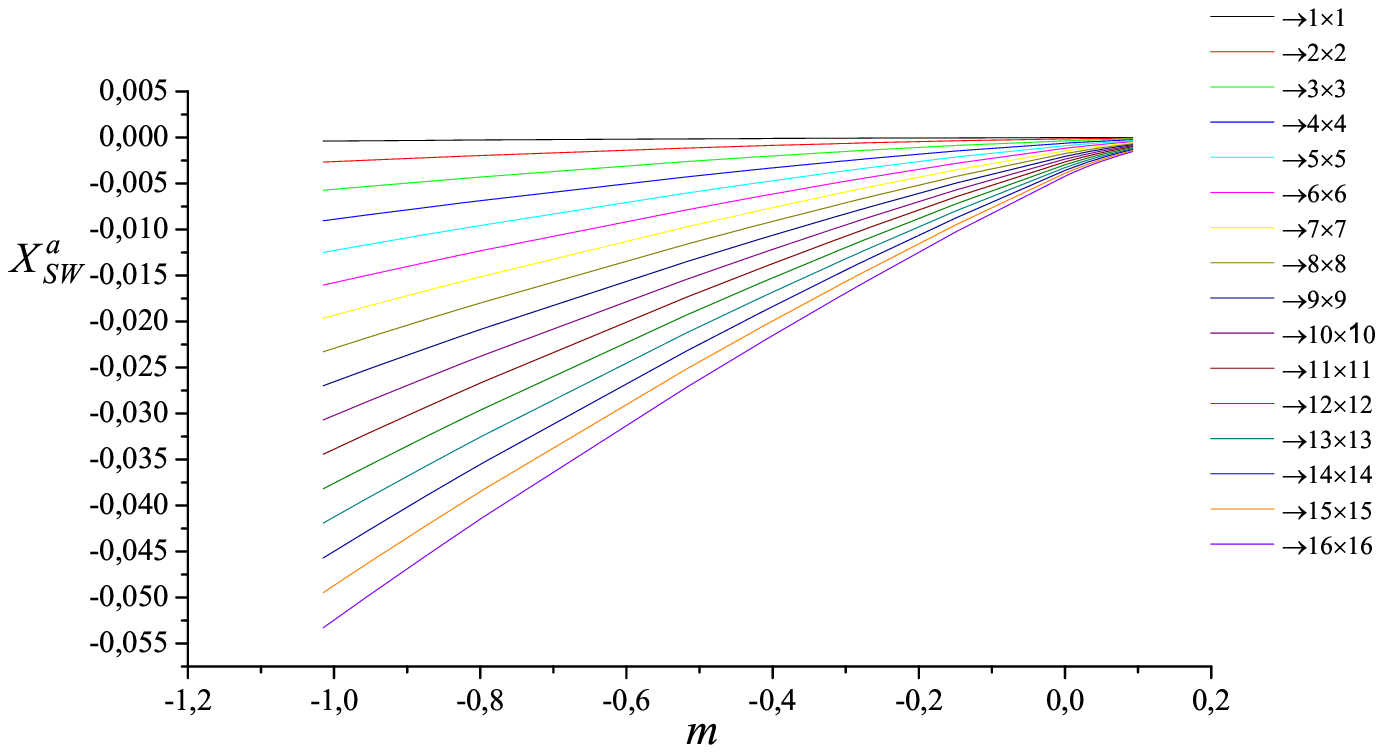}

{\footnotesize Fig.3: $X^a_{SW}$ for loops $L\times L$. Top line: $L=1$, bottom line: $L=16$.}
\end{center}

\begin{center}
\psfig{width=15truecm,file=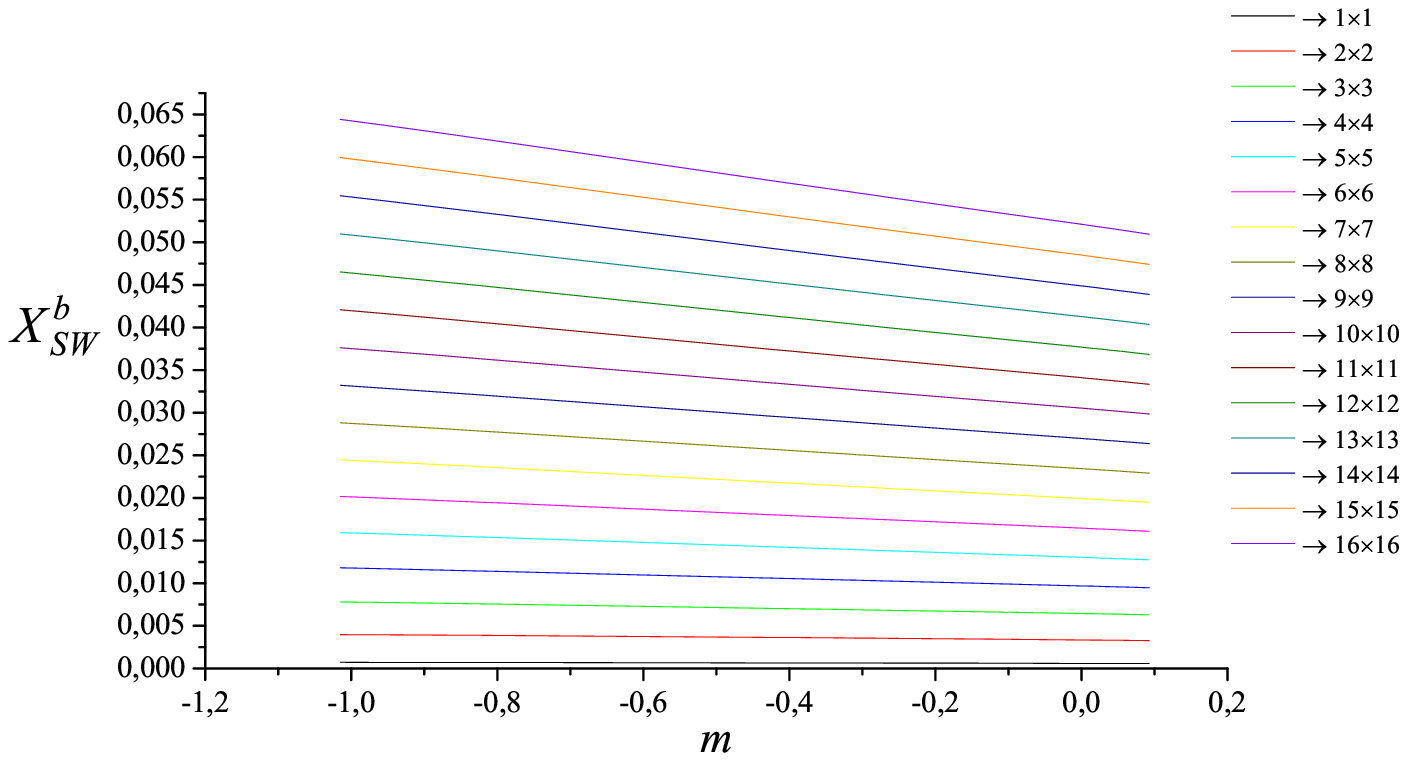}

{\footnotesize Fig.4: $X^b_{SW}$ for loops $L\times L$. Top line: $L=16$, bottom line: $L=1$.} 
\end{center}

\eject
\section{Calculation with Overlap Fermions}
\label{Calculation with Overlap Fermions}

The fermionic action now reads 

\begin{equation}
S_f =\sum_f \sum_{x,y} \bar{\psi}^f_x D_{\rm N}\left( x,y \right )\psi^f_y.
\label{latact}
\end{equation}

\noindent
with: $D_{\rm N} = M_O \left[1 + X (X^\dagger X)^{-1/2} \right]$,
and: $X = D_{\rm W} - M_O$. Here, $D_{\rm W}$ is the massless
Wilson-Dirac operator with $r=1$, and $M_O$ is a free parameter whose
value must be in the range $0 <M_O < 2$, in order to guarantee the correct 
pole structure of $D_{\rm N}$.

Fermionic vertices are obtained by separating the Fourier transform 
of $D_{\rm N}$ into a free part (inverse propagator $D_0$) and an interaction 
part $\Sigma$.

\begin{equation}
(1/M_O) D_{\rm N}(q,p)=D_0 (p) (2\pi)^4\delta^4(q-p) + \Sigma(q,p),
\end{equation}

\noindent
where:

\begin{equation}
D_0^{-1}(p) = {-i\sum_\mu \gamma_\mu \sin p_\mu \over 2 \left[ \omega(p) + b(p)\right] } + {1\over 2},
\label{d0}
\end{equation}
\begin{eqnarray}
{\rm and:}\qquad \omega(p) &=& \left( \sum_\mu \sin^2 p_\mu + \bigl[ \sum_\mu (1-\cos p_\mu ) - M_O \bigr]^2 \right)^{1/2}, \\
b(p)&=& \sum_\mu (1-\cos p_\mu) -  M_O.
\end{eqnarray}

$\Sigma\left(q,p\right)$ is given by:
\begin{equation}
\Sigma(q,p) = V^1 + V_1^2 + V_2^2 + {\cal{O}}(g^3),
\end{equation}
\noindent
with:
\begin{eqnarray}
V^1 &=& {1\over \omega(p) + \omega(q)}
\left[X_1(q,p) - {1\over \omega(p)\omega(q)} X_0(q) X^\dagger_1(q,p) X_0(p)\right] \nonumber\\
V_1^2 &=& {1\over \omega(p) + \omega(q)}
\left[X_2(q,p) - {1\over \omega(p)\omega(q)} X_0(q) X^\dagger_2(q,p)
X_0(p)\right] \nonumber\\
V_2^2 &=& \int {d^4 k\over (2\pi)^4}
{1\over \omega(p) + \omega(q)}\ 
{1\over \omega(p) + \omega(k)}\ 
{1\over \omega(q) + \omega(k)}\nonumber\\
&\times&
\!\!\Biggl[
-X_0(q)X_1^\dagger(q,k)X_1(k,p) 
 -X_1(q,k)X_0^\dagger(k)X_1(k,p)\nonumber\\ &-&X_1(q,k)X_1^\dagger(k,p)X_0(p) 
\;\;+{\omega(p)+\omega(q)+\omega(k)\over \omega(p)\omega(q)\omega(k)}\nonumber\\
&\times&X_0(q)X_1^\dagger(q,k)X_0(k)X_1^\dagger(k,p)X_0(p)\Biggr] + {\cal{O}}(g^3).\nonumber
\end{eqnarray}

\noindent
$X_0,X_1$ and $X_2$ denote the parts of the Dirac-Wilson operator with 0, 1 and  2 
gluons (of order ${\cal{O}}(g^0)$,${\cal{O}}(g^1)$ and ${\cal{O}}(g^2)$).

The two different types of contributions, $V_1^2$ and $V_2^2$, which comprise the 2-gluon vertex may be depicted as in Figs. 5 and 6, respectively.

\begin{center}
\begin{minipage}{0.4\linewidth}
\begin{center}
\psfig{figure=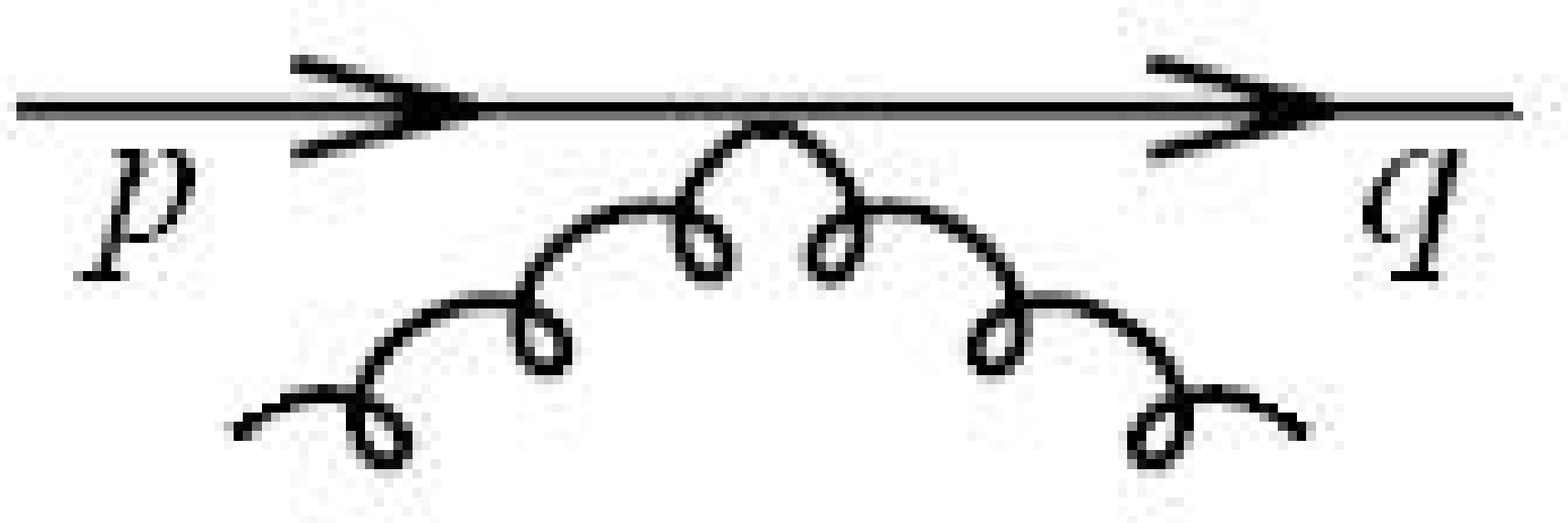,height= 2truecm}\\
\end{center}
%\vskip-2.7cm \hskip1.7cm $\vec p$\hskip2.8cm $\vec q$\vskip2.7cm
%\vskip-0.5cm
\hskip0.8cm{\footnotesize Fig.5: Pointlike part of the vertex.\\
\phantom{Some more space here...}}
\end{minipage} \hskip0.1\textwidth
\begin{minipage}{0.4\linewidth}
\begin{center}
\psfig{figure=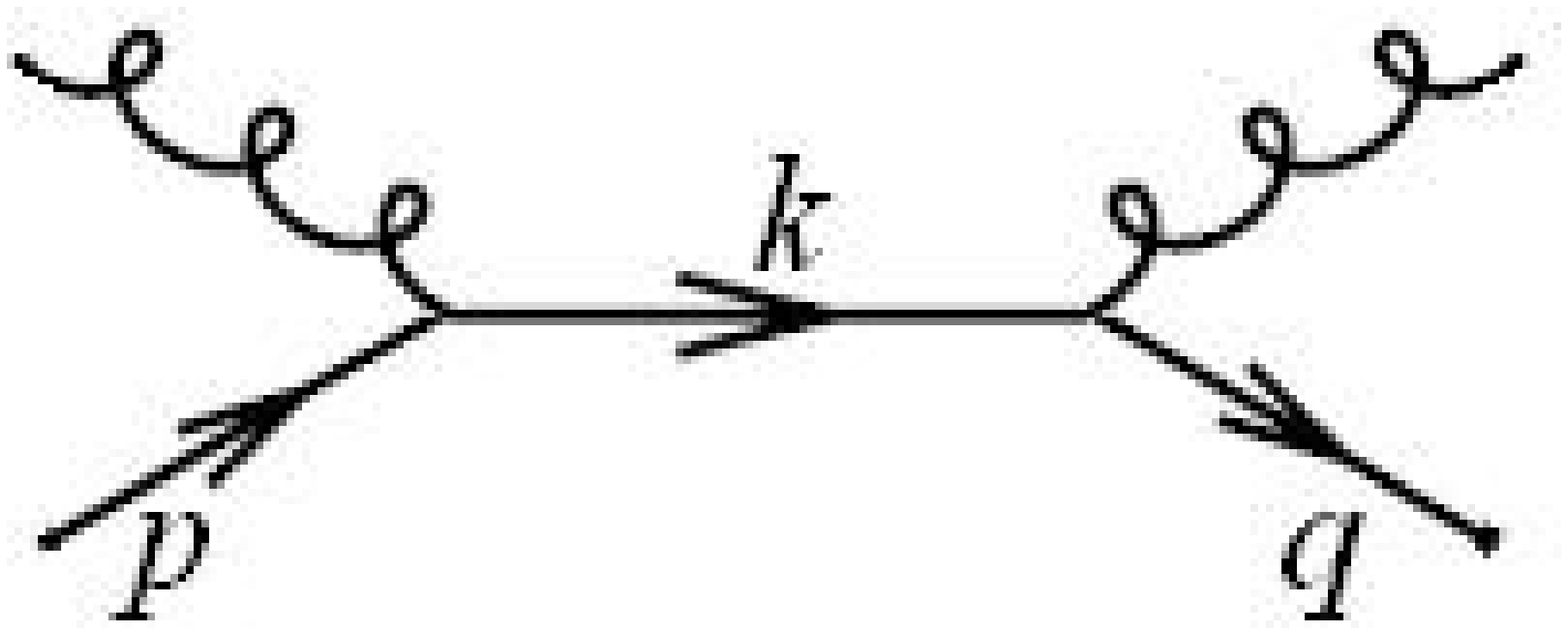,height= 2truecm}\\
\end{center}
%\vskip-1cm \hskip0.8cm $\vec p$\hskip4.5cm $\vec q$ \vskip1cm
%\vskip-2.8cm \hskip3.3cm $\vec k$ \vskip2.8cm
%\vskip-1cm
{\footnotesize Fig.6: Vertex part with intermediate momentum
integration.}
\end{minipage}
\end{center}
\medskip

The perturbative expansion of the Wilson loop is given by the expression:

\begin{equation}
W\left(C\right)\, /\, N=1-W_{LO}g^2-\left(W_{NLO}-{\left(N^2-1\right)\over N}N_f\Mulberry{X_{Overlap}}\right)g^4,
\label{Xoverlap}
\end{equation}

\noindent
where $\Mulberry{X_{Overlap}}$ are the values we compute. Fig. 7 shows $\Mulberry{X_{Overlap}}$ as a function of $M_O$ for square Wilson loops. Numerical values are presented in Table \ref{Overlap}.

\begin{center}
\psfig{width=15truecm,file=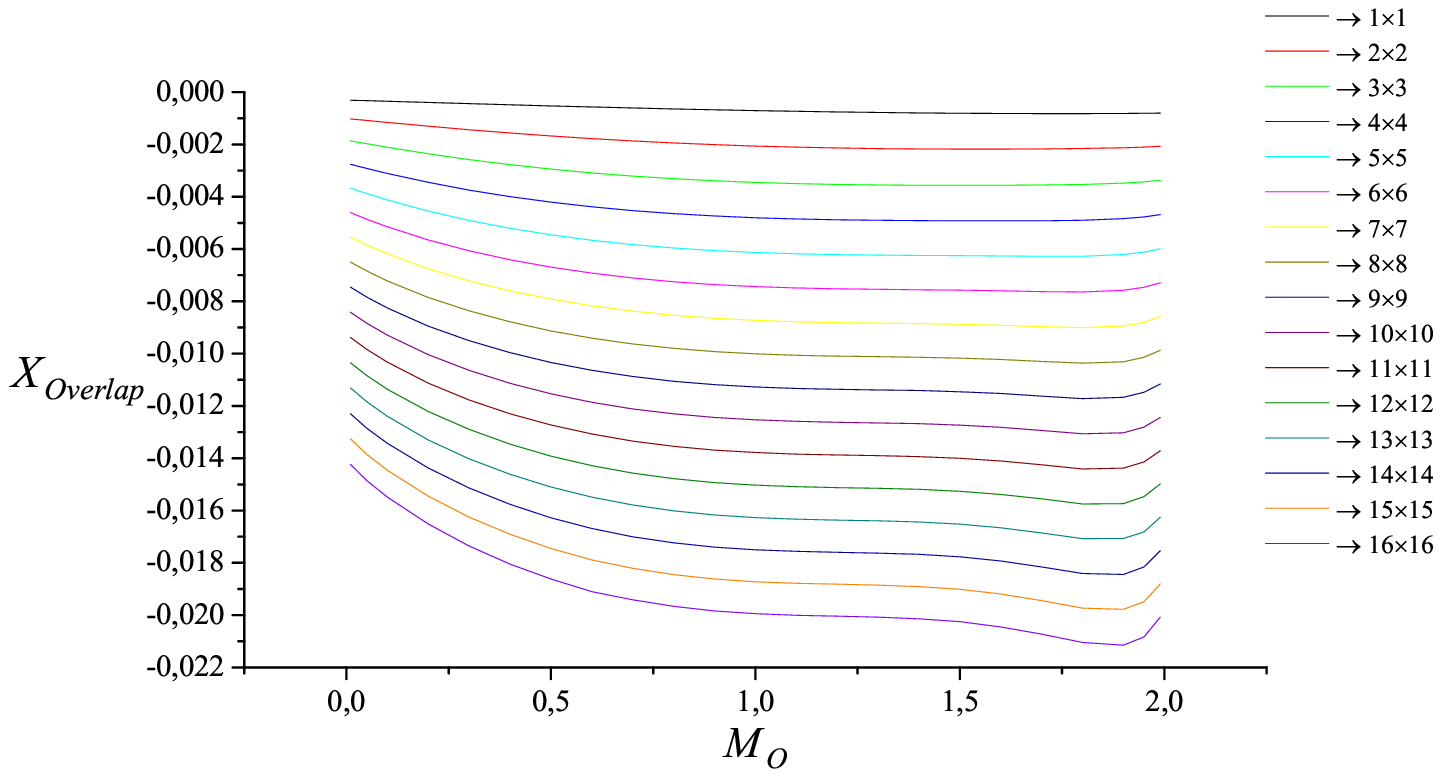}

{\footnotesize Fig.7: $X_{Overlap}$ for loops $L\times L$. Top line: $L=1$, bottom line: $L=16$.}
\end{center}

\eject
\section{Calculation of the b-quark mass shift}
\label{Calculation of the b-quark mass shift}

In perturbation theory, the expectation value of large Wilson loops decreases exponentially with the perimeter of the loops:

\begin{equation}
\left\langle W\left(R,T\right) \right\rangle \sim {\rm exp}\left(-2\delta m\left(R+T\right)\right).
\end{equation}

\noindent
Following Ref.~\cite{GMCTS}, the perturbative expansion for $\left\langle W\left(R,T\right) \right\rangle$ is:

\begin{equation}
\left\langle W\left(R,T\right) \right\rangle = 1 - g^2\, W_2\left(R,T\right)- g^4\, \Brown{W_4\left(R,T\right)}+ {\cal O}\left(g^6\right).
\end{equation}

\noindent
Using the expectation value of $W\left(R,T\right)$, we obtain the perturbative expansion for $\delta m$:

\begin{eqnarray}
\delta m = {1\over 2\left(R+T\right)}\Bigg[g^2\, W_2\left(R,T\right)+ g^4\left(\Brown{W_4\left(R,T\right)}+{1\over2}W^2_2\left(R,T\right)\right)\Bigg],
\end{eqnarray}

\noindent
where: $W_2\left(R,T\right)$ involves only gluons and: 

\begin{equation}
\Brown{W_4\left(R,T\right)}=\Brown{W^{g}_4\left(R,T\right)}+\Brown{W^{f}_4\left(R,T\right)}.
\end{equation}

\noindent
$\Brown{W^{g}_4\left(R,T\right)}$ is the contribution in the pure-gauge theory and $\Brown{W^{f}_4\left(R,T\right)}$ is the fermionic contribution. To evaluate the effect of fermions on the mass shift, we must examine their contribution in the limit as $R, T \rightarrow \infty$. To this end, we note that our expression assumes the generic form (modulo terms which will not contribute in this limit):

\begin{eqnarray}
\int{d^4p\over \left(2\pi\right)^4} \sin^2\left({p_{\nu}T\over2}\right)\sin^2\left({p_{\mu}R\over2}\right)\left({1\over \sin^2\left({p_{\nu}/ 2}\right)} +{1\over \sin^2\left({p_{\mu}/ 2}\right)}\right){1\over\left(\hat{p}^2\right)^2}G\left(p\right),
\label{Int}
\end{eqnarray}

\noindent
where $\hat{p}^2=4\displaystyle{\sum_{\rho}} \sin^2\left(p_{\rho}/2\right)$ and $G\left(p\right)\sim p^2$. As $R, T \rightarrow \infty$, the above expression becomes:

\begin{eqnarray}
{1\over2}\left(R+T\right)\int {d^3\bar{p}\over \left(2\pi\right)^3}{1\over\left(\hat{\bar{p}}^2\right)^2}G\left(\bar{p}\right),
\end{eqnarray}

\noindent
where: $\bar{p}=\left(p_1,p_2,p_3,0\right)$ (for $\mu=4$ or $\nu=4$). 

For the clover action, the fermionic contribution takes the form:

\begin{eqnarray}
\Brown{W^{f}_4\left(R,T\right)} = \left(N^2-1\right)N_f\left(R+T\right)\left(\Brown{V_W}+\Brown{V^a_{SW}}\Blue{c_{SW}}+\Brown{V^b_{SW}}\Blue{c^2_{SW}}\right),
\end{eqnarray}

\noindent
and for the overlap action, it reads:
\begin{equation}
\Brown{W^{f}_4\left(R,T\right)} = \left(N^2-1\right)N_f\left(R+T\right)\Brown{V_{Overlap}}.
\end{equation}

\noindent
The values of \Brown{$V_W$}, \Brown{$V^a_{SW}$}, \Brown{$V^b_{SW}$} and \Brown{$V_{Overlap}$} have been calculated in the present work. Figs.~8,~9 show their values as a function of the bare fermion mass for clover fermions and of $M_O$ for overlap fermions:

\begin{center}
  \psfig{width=15truecm,file=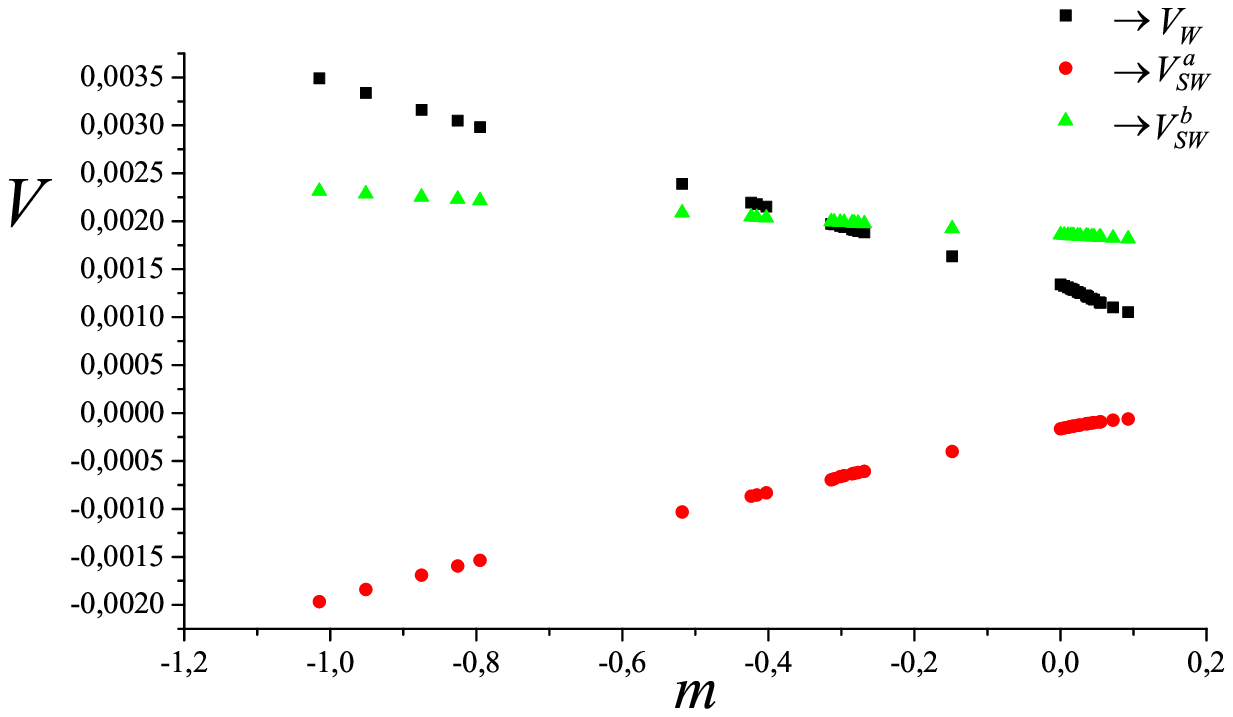}

{\footnotesize Fig.8: $V_W$, $V^a_{SW}$ and $V^b_{SW}$ as a function of $m$.} 
\end{center}

\begin{center}
  \psfig{width=15truecm,file=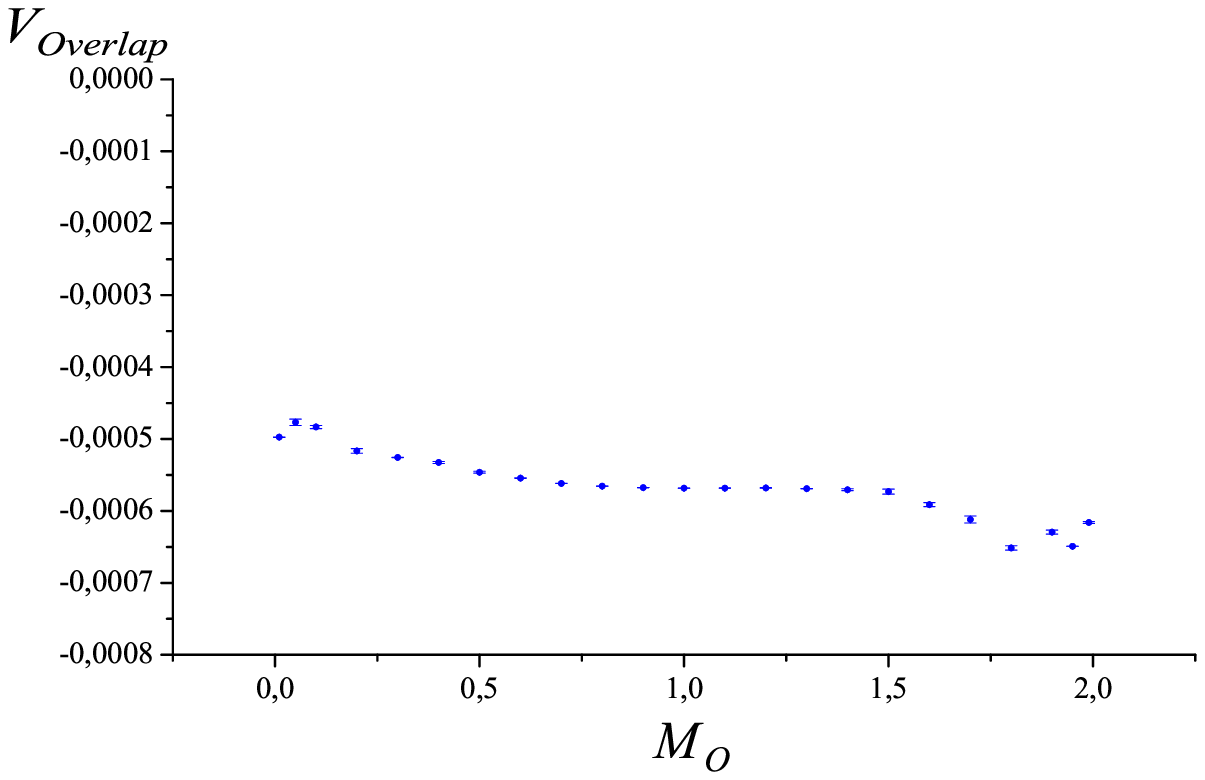}

{\footnotesize Fig.9: $V_{Overlap}$ as a function of $M_O$.} 
\end{center}

\eject
\noindent
At \MidnightBlue{one-loop} order the b-quark mass shift is given by:

\begin{equation}
a\delta m \simeq 0.16849 g^2 + {\cal{O}}\left( g^4\right),
\end{equation}

\noindent
if we set the number of colours N equal to 3. 

Using the results for the infinite spatial and temporal extent we arrive at the \MidnightBlue{{\bf two-loop}} expression for $\delta m$. We list below some examples:

\noindent
\NavyBlue{$\bullet$} The general form of $\delta m$ for \MidnightBlue{clover} fermions is ($\alpha_0=g^2/4\pi$):

\begin{eqnarray}
a \delta m \simeq 2.1173\alpha_0&+&\Bigg[11.152 - {\left(4\pi\right)^2\left(N^2-1\right)N_f \over 2 N}\nonumber\\
&\times&\left(\Brown{V_W}+\Brown{V_{SW}^a}\Blue{c_{SW}}+\Brown{V_{SW}^b} \Blue{c_{SW}^2}\right)\Bigg]\alpha_0^2+{\cal{O}}\left(\alpha_0^3\right).
\label{shiftclover}
\end{eqnarray}

\noindent
The values of \Brown{$V_W(m)$}, \Brown{$V_{SW}^a(m)$}, \Brown{$V_{SW}^b(m)$} are listed in Table~\ref{InfClover}. In particular, setting $m=0.0$:

\begin{eqnarray}
a \delta m \simeq 2.1173\alpha_0&+&\Bigg[11.152 - {\left(4\pi\right)^2\left(N^2-1\right)N_f \over 2 N}\Bigg(\Red{0.00134096(5)}\nonumber \\
\Red{&-& 0.0001641(1)}\, \Blue{c_{SW}}+\Red{0.00185871(2)}\, \Blue{c_{SW}^2}\Bigg)\Bigg]\alpha_0^2+{\cal{O}}\left(\alpha_0^3\right).
\end{eqnarray}

\noindent
These numbers agree with Ref.~\cite{GMCTS}, within the precision presented there.

\noindent
\NavyBlue{$\bullet$} For \MidnightBlue{overlap} fermions, the general form of $\delta m$ is:

\begin{eqnarray}
a \delta m \simeq 2.1173\alpha_0+\Bigg[11.152 - {\left(4\pi\right)^2\left(N^2-1\right)N_f \over 2 N}\Brown{V_{Overlap}}\Bigg]\alpha_0^2+{\cal{O}}\left(\alpha_0^3\right).
\label{shiftoverlap}
\end{eqnarray}

\noindent
The values of \Brown{$V_{Overlap}$}, are listed in Table~\ref{InfOverlap}. In particular, setting $M_O=1.4$:

\begin{eqnarray}
a \delta m \simeq 2.1173\alpha_0+\Bigg[11.152 - {\left(4\pi\right)^2\left(N^2-1\right)N_f \over 2 N}
\Red{0.000571(1)}\Bigg]\alpha_0^2+{\cal{O}}\left(\alpha_0^3\right).
\end{eqnarray}

\section{Calculation of the Perturbative Static Potential}
\label{Calculation of the Perturbative Static Potential}

The static potential is given by the expression:

\begin{eqnarray} 
aV\left(Ra\right) &=& -\lim_{T \rightarrow \infty} {d \ln W\left(R,T\right)\over dT}\\
&=& V_1\left(R\right)g^2+V_2\left(R\right)g^4+\ldots,
\end{eqnarray}

\noindent
where: $V_1\left(R\right)$ is a pure gluonic contribution ~\cite{GBPB}. $V_2\left(R\right)$ contains a gluonic part $V_g \left(R\right)$, which can be found in Ref.~\cite{GBPB},
and a fermionic part \PineGreen{$F\left(R\right)$}:

\begin{eqnarray}
V_2\left(R\right) &=& V_g\left(R\right) -{\left(N^2-1\right)\over N}N_f\PineGreen{F\left(R\right)}.
\label{pot}
\end{eqnarray}

\noindent
We compute \PineGreen{$F\left(R\right)$} for clover and overlap fermions. For \MidnightBlue{clover}:

\begin{eqnarray}
\PineGreen{F\left(R\right)} = \PineGreen{F_{Clover}\left(R\right)}= F_W\left(R\right)+ F^a_{SW}\left(R\right)\Blue{c_{SW}}+ F^b_{SW}\left(R\right)\Blue{c^2_{SW}}
\end{eqnarray}

\noindent
and for \MidnightBlue{overlap}: \PineGreen{$F\left(R\right)} =\PineGreen{F_{Overlap}\left(R\right)$}. The values of $F_{Clover}\left(R\right)$ ($\Blue{c_{SW}}=1.3$) and $F_{Overlap}\left(R\right)$ for specific mass values, can be found in Tables \ref{CloverPotential} and \ref{OverlapPotential}. We present them in Figs.~10-11 as a function of $R$. The values of $F_W$, $F^a_{SW}$ and $F^b_{SW}$ as a function of $m$ are shown in Figs.~12-14, and they can be found in Tables \ref{CloverVSw0} - \ref{CloverVSw2}.

\begin{center}
  \psfig{width=15truecm,file=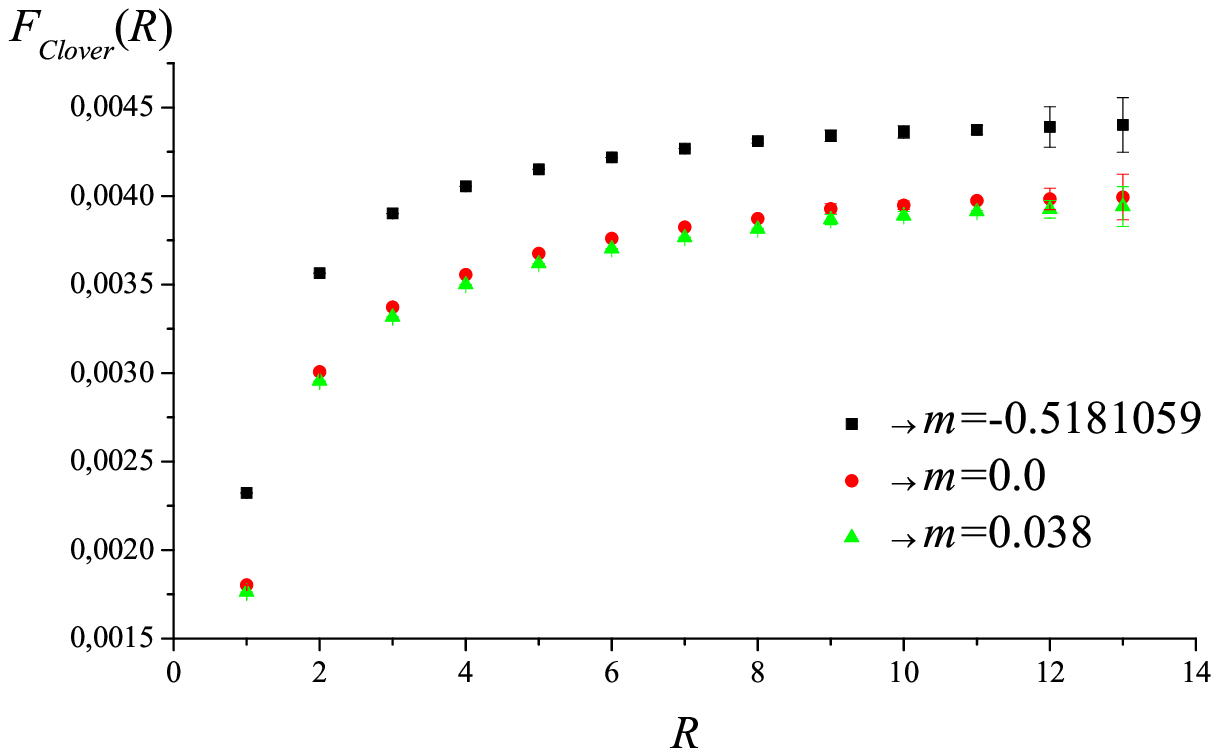}

{\footnotesize Fig.10: $F_{Clover}$ as a function of $R$ ($c_{SW}$ = 1.3).}
\end{center} 

\eject
\begin{center}
  \psfig{width=15truecm,file=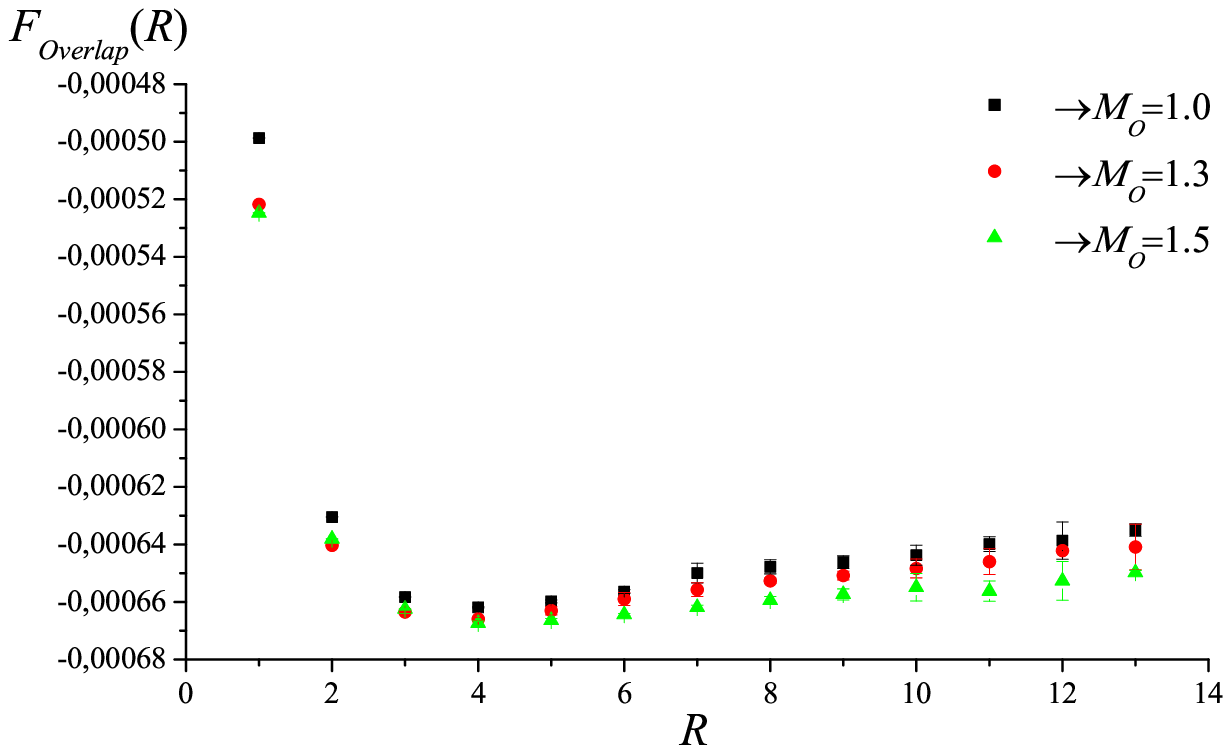}

{\footnotesize Fig.11: $F_{Overlap}$ as a function of $R$.}
\end{center} 

\begin{center}
  \psfig{width=15truecm,file=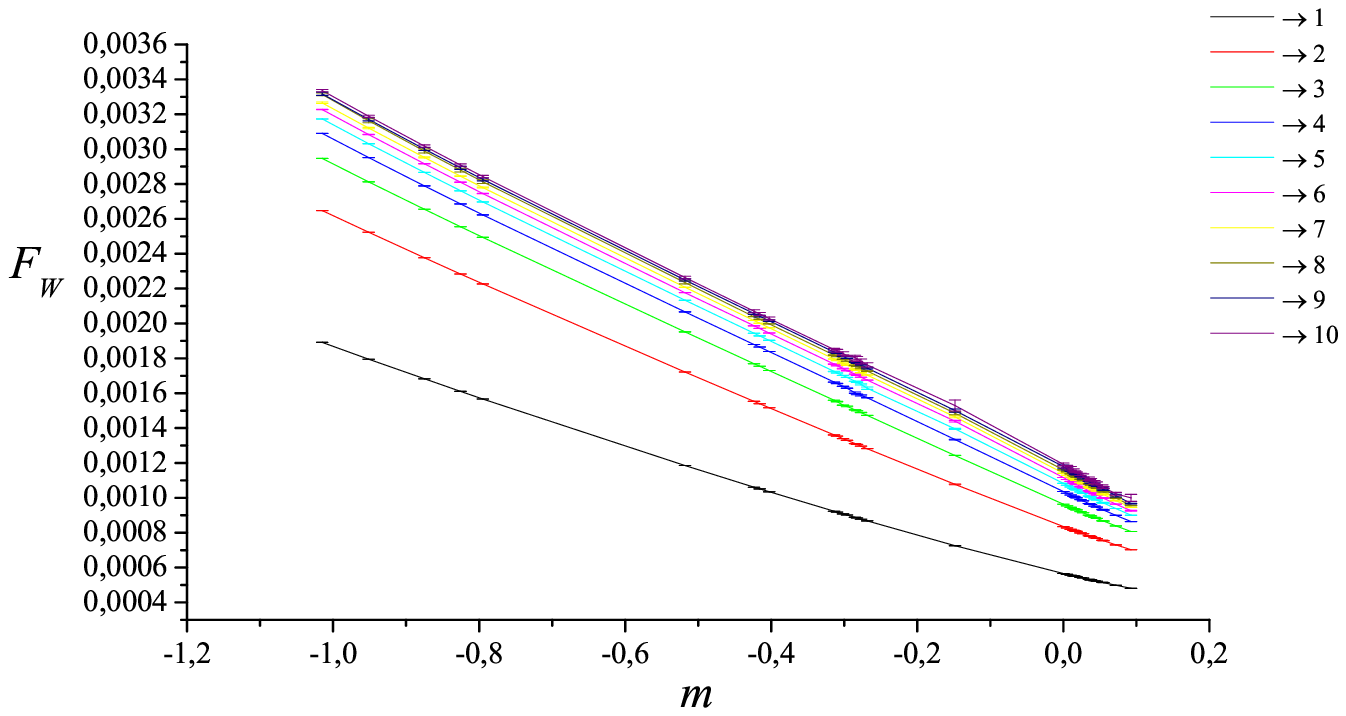}

{\footnotesize Fig.12: $F_{W}$ as a function of the bare fermion mass $m$. Top line: $R=10$, bottom line: $R=1$.}
\end{center} 
\begin{center}
  \psfig{width=15truecm,file=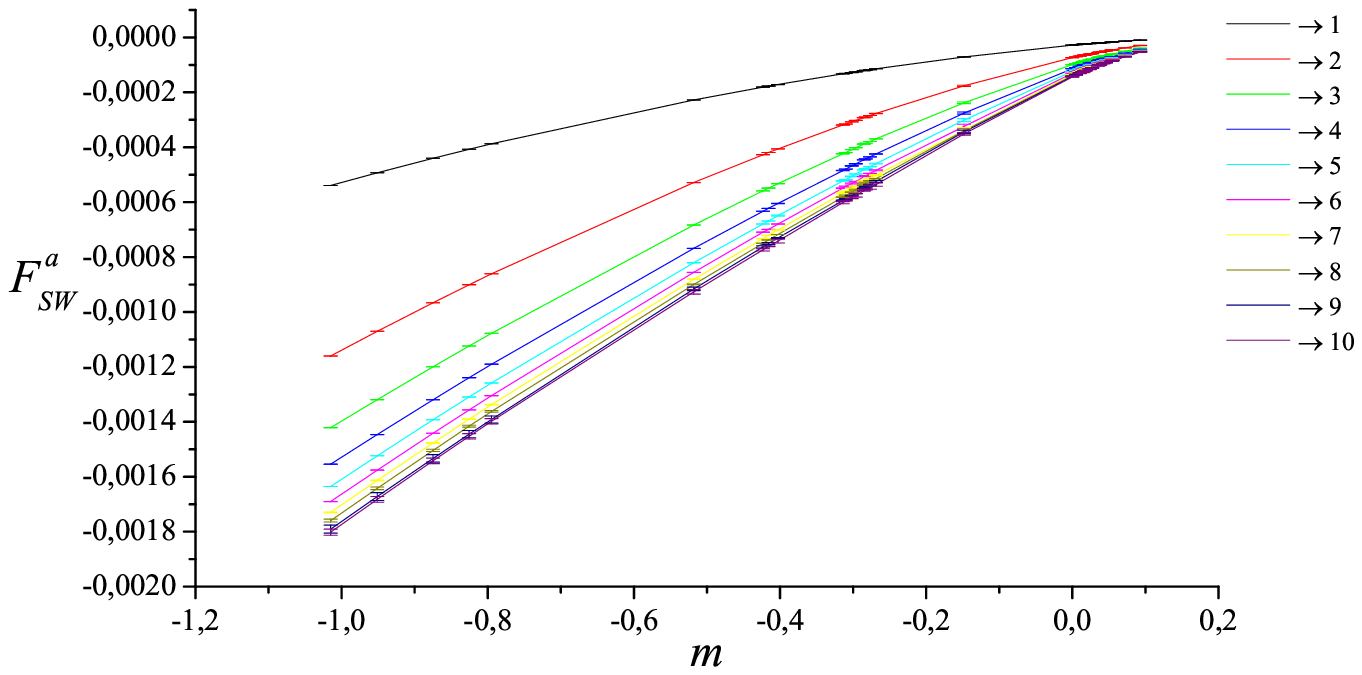}

{\footnotesize Fig.13: $F^a_{SW}$ as a function of the bare fermion mass $m$. Top line: $R=1$, bottom line: $R=10$.}
\end{center} 

\begin{center}
  \psfig{width=15truecm,file=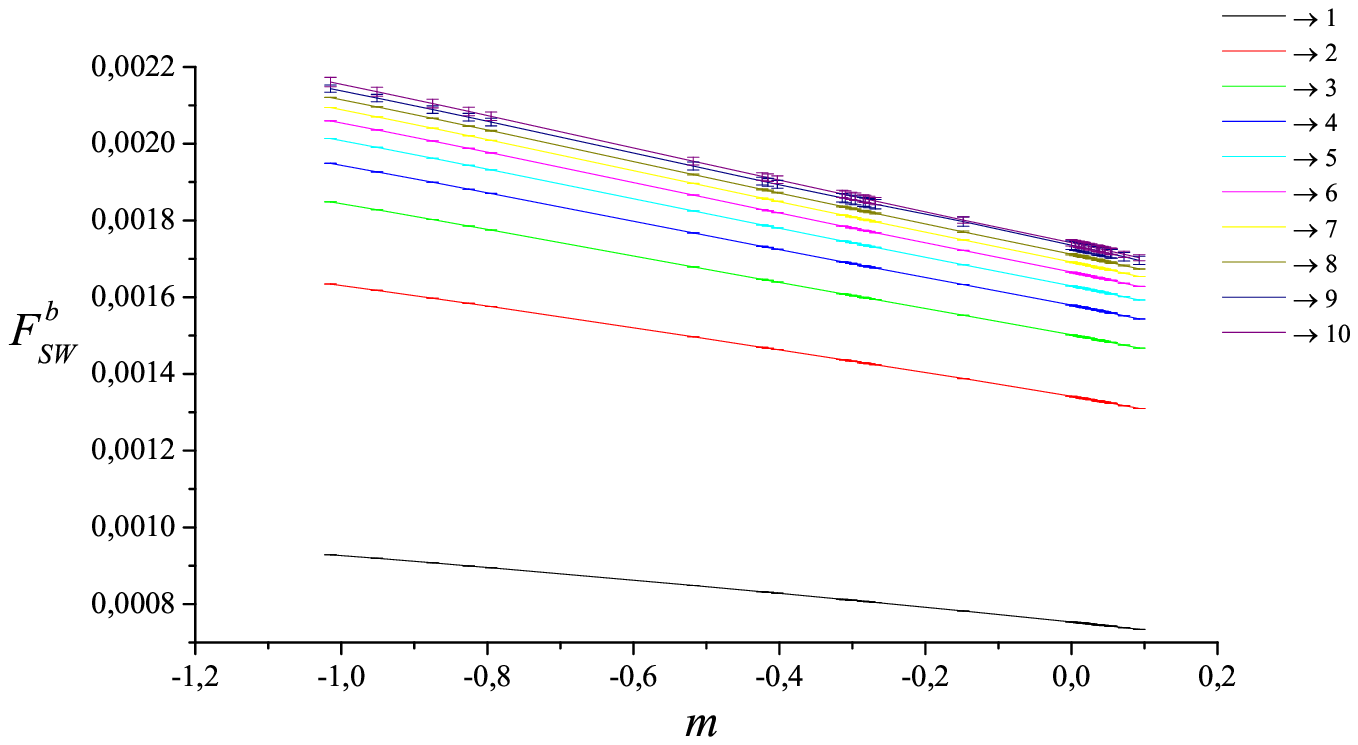}

{\footnotesize Fig.14: $F^b_{SW}$ as a function of the bare fermion mass $m$. Top line: $R=10$, bottom line: $R=1$.}
\end{center}

\begin{table}[ht]
\begin{center}
\begin{minipage}{17cm}
\caption{Values of \Red{{\bf $X_{W}$}}, Eqs.(\ref{Wc}, \ref{Xi}), for specific masses.
\label{CloverSw0}}
\begin{tabular}{r@{}lr@{}lr@{}lr@{}lr@{}lr@{}l}
\multicolumn{2}{c}{$R\times T$}&
\multicolumn{2}{c}{$m=-0.8253968$} &
\multicolumn{2}{c}{$m=-0.5181059$} &
\multicolumn{2}{c}{$m=-0.1482168$} &
\multicolumn{2}{c}{$m=0.0$} &
\multicolumn{2}{c}{$m=0.038$} \\
\tableline \hline

  %M*N           m=-0.8253968        m=-0.5181059      m=-0.1482168        m=0.0             m=0.038
\\
1&$\times$1 &0&.002017985941(3) &0&.0014485866(6) &0&.000877986(6) &0&.0006929202(2) &0&.000651020(4)\\
2&$\times$1 &0&.003808789532(5) &0&.0027649262(6) &0&.001680527(8) &0&.0013187555(4) &0&.001236379(7)\\
2&$\times$2 &0&.006695252630(2) &0&.0049398240(9) &0&.00302703(1)  &0&.0023609937(7) &0&.00220774(2)\\
3&$\times$1 &0&.0054740966(1)   &0&.0039933034(4) &0&.002433213(9) &0&.0019053135(4) &0&.001784665(8)\\
3&$\times$2 &0&.0091696433(5)   &0&.006816343(1)  &0&.00420265(3)  &0&.003270793(4)  &0&.00305476(4)\\
3&$\times$3 &0&.012099698(1)    &0&.009076160(3)  &0&.0056468(1)   &0&.004385778(7)  &0&.00409027(4)\\
4&$\times$1 &0&.0071034996(5)   &0&.0051941492(2) &0&.003169672(9) &0&.0024795565(8) &0&.00232145(2)\\
4&$\times$2 &0&.011522152(2)    &0&.008597002(3)  &0&.00532068(8)  &0&.0041372190(9) &0&.00386148(4)\\
4&$\times$3 &0&.014795778(3)    &0&.011149549(3)  &0&.0069774(2)   &0&.005415529(4)  &0&.00504665(5)\\
4&$\times$4 &0&.017710525(6)    &0&.013413919(4)  &0&.0084565(3)   &0&.006560910(5)  &0&.0061084(2)\\
5&$\times$1 &0&.008722165(1)    &0&.006386130(1)  &0&.00390045(1)  &0&.003049571(1)  &0&.00285433(3)\\
5&$\times$2 &0&.013836216(4)    &0&.010345426(2)  &0&.0064176(2)   &0&.004988015(3)  &0&.00465378(6)\\
5&$\times$3 &0&.017414883(7)    &0&.013157873(3)  &0&.0082648(3)   &0&.006413278(5)  &0&.0059734(2)\\
5&$\times$4 &0&.02050333(1)     &0&.015574716(3)  &0&.0098662(4)   &0&.007654746(6)  &0&.0071224(3)\\
5&$\times$5 &0&.02342108(2)     &0&.017846119(6)  &0&.0113705(9)   &0&.00882313(1)   &0&.0082041(7)\\
6&$\times$1 &0&.010336971(2)    &0&.0075747735(4) &0&.00462888(3)  &0&.003617837(3)  &0&.00338559(4)\\
6&$\times$2 &0&.016135852(8)    &0&.0120812913(6) &0&.0075056(2)   &0&.005832123(6)  &0&.00543991(5)\\
6&$\times$3 &0&.02000402(2)     &0&.0151400270(8) &0&.0095331(3)   &0&.007396912(8)  &0&.0068870(2)\\
6&$\times$4 &0&.02324722(2)     &0&.017692706(1)  &0&.0112452(9)   &0&.008725234(9)  &0&.0081151(3)\\
6&$\times$5 &0&.02626871(3)     &0&.020056063(2)  &0&.012828(1)    &0&.00995767(2)   &0&.0092550(4)\\
6&$\times$6 &0&.02919736(5)     &0&.02233797(1)   &0&.014349(1)    &0&.01114489(4)   &0&.0103532(6)\\
7&$\times$1 &0&.011950092(5)    &0&.0087619313(1) &0&.00535615(5)  &0&.004185240(5)  &0&.00391607(5)\\
7&$\times$2 &0&.01842902(2)     &0&.013811440(5)  &0&.0085891(3)   &0&.006672898(9)  &0&.0062228(1)\\
7&$\times$3 &0&.02257938(3)     &0&.01710999(1)   &0&.0107918(3)   &0&.00837337(2)   &0&.0077942(3)\\
7&$\times$4 &0&.02596811(5)     &0&.01979032(2)   &0&.0126090(6)   &0&.00978362(2)   &0&.0090969(5)\\
7&$\times$5 &0&.02908270(7)     &0&.02223621(3)   &0&.0142608(8)   &0&.01107437(3)   &0&.0102893(7)\\
7&$\times$6 &0&.0320807(1)      &0&.02457977(4)   &0&.015838(1)    &0&.01230793(4)   &0&.0114291(8)\\
7&$\times$7 &0&.0350212(2)      &0&.02687227(8)   &0&.017374(2)    &0&.01351051(4)   &0&.0125406(8)\\
8&$\times$1 &0&.01356236(1)     &0&.009948329(4)  &0&.00608280(8)  &0&.00475217(1)   &0&.00444608(5)\\
8&$\times$2 &0&.02071885(4)     &0&.01553863(2)   &0&.0096702(1)   &0&.00751181(4)   &0&.0070043(2)\\
8&$\times$3 &0&.02514752(8)     &0&.01907354(5)   &0&.0120453(2)   &0&.00934576(4)   &0&.0086978(5)\\
8&$\times$4 &0&.0286767(1)      &0&.02187717(8)   &0&.0139600(6)   &0&.01083507(7)   &0&.0100722(7)\\
8&$\times$5 &0&.0318785(2)      &0&.0244002(1)    &0&.015681(1)    &0&.01218072(7)   &0&.0113140(8)\\
8&$\times$6 &0&.0349394(2)      &0&.0267996(1)    &0&.017309(1)    &0&.01345678(7)   &0&.0124921(8)\\
8&$\times$7 &0&.0379297(3)      &0&.0291364(2)    &0&.018886(1)    &0&.01469473(8)   &0&.0136356(9)\\
8&$\times$8 &0&.0408807(6)      &0&.0314384(4)    &0&.020433(1)    &0&.0159099(2)    &0&.014758(2)\\

\end{tabular}
\end{minipage}
\end{center}
\end{table}

\begin{table}[ht]
%\linespread 1
\begin{center}
\begin{minipage}{17cm}
\caption{Values of \Red{{\bf $X^{a}_{SW}$}}, Eqs.(\ref{Wc}, \ref{Xi}), for specific masses.
\label{CloverSw1}}
\begin{tabular}{r@{}lr@{}lr@{}lr@{}lr@{}lr@{}l}
\multicolumn{2}{c}{$R\times T$}&
\multicolumn{2}{c}{$m=-0.8253968$} &
\multicolumn{2}{c}{$m=-0.5181059$} &
\multicolumn{2}{c}{$m=-0.1482168$} &
\multicolumn{2}{c}{$m=0.0$} &
\multicolumn{2}{c}{$m=0.038$} \\
\tableline \hline

\\
1&$\times$1 &-0&.000292493808(2) &-0&.0001573984(1)  &-0&.00004774(1)  &-0&.00002010061(1)  &-0&.000014684(1)\\
2&$\times$1 &-0&.000789014304(3) &-0&.0004331752(1)  &-0&.00013418(1)  &-0&.0000570082(3)   &-0&.000041862(3)\\
2&$\times$2 &-0&.00204528413(5)  &-0&.0011551861(2)  &-0&.00037017(3)  &-0&.0001607273(9)   &-0&.000119471(9)\\
3&$\times$1 &-0&.001229832201(3) &-0&.0006828156(2)  &-0&.00021266(2)  &-0&.0000890331(1)   &-0&.000064740(3)\\
3&$\times$2 &-0&.0030768902(1)   &-0&.0017689377(8)  &-0&.00057553(4)  &-0&.0002481672(1)   &-0&.00018354(2)\\
3&$\times$3 &-0&.0044714337(3)   &-0&.002630731(5)   &-0&.00087552(5)  &-0&.0003758198(2)   &-0&.00027679(2)\\
4&$\times$1 &-0&.001650286801(4) &-0&.0009202841(3)  &-0&.00028722(2)  &-0&.0001190532(1)   &-0&.000085986(7)\\
4&$\times$2 &-0&.0040293806(2)   &-0&.002335182(4)   &-0&.00076609(5)  &-0&.0003285984(3)   &-0&.00024206(2)\\
4&$\times$3 &-0&.0057054610(5)   &-0&.003394523(5)   &-0&.00114482(6)  &-0&.0004893900(3)   &-0&.00035909(2)\\
4&$\times$4 &-0&.007128162(1)    &-0&.004297068(7)   &-0&.00147396(7)  &-0&.0006277562(3)   &-0&.00045880(6)\\
5&$\times$1 &-0&.0020638727(1)   &-0&.0011528916(8)  &-0&.00035998(2)  &-0&.0001481833(3)   &-0&.000106510(8)\\
5&$\times$2 &-0&.0049550067(3)   &-0&.002882530(4)   &-0&.00094971(5)  &-0&.0004057803(5)   &-0&.00029804(1)\\
5&$\times$3 &-0&.0068832569(7)   &-0&.004118532(9)   &-0&.00139950(7)  &-0&.0005962550(5)   &-0&.00043618(5)\\
5&$\times$4 &-0&.0084593585(7)   &-0&.00513455(1)    &-0&.00177899(7)  &-0&.0007552273(5)   &-0&.00055013(8)\\
5&$\times$5 &-0&.0099050827(9)   &-0&.00605756(2)    &-0&.00212377(6)  &-0&.0008986526(5)   &-0&.0006522(1)\\
6&$\times$1 &-0&.0024746461(6)   &-0&.001383364(1)   &-0&.00043184(2)  &-0&.0001768954(5)   &-0&.000126713(9)\\
6&$\times$2 &-0&.005869585(2)    &-0&.003421541(5)   &-0&.00112989(5)  &-0&.0004813887(8)   &-0&.00035280(4)\\
6&$\times$3 &-0&.008037374(3)    &-0&.00482471(1)    &-0&.00164679(6)  &-0&.000699807(1)    &-0&.00051074(7)\\
6&$\times$4 &-0&.009751051(4)    &-0&.00594233(2)    &-0&.00207166(6)  &-0&.000877229(4)    &-0&.0006373(1)\\
6&$\times$5 &-0&.011293280(5)    &-0&.00693732(2)    &-0&.00245048(4)  &-0&.001034160(4)    &-0&.0007484(1)\\
6&$\times$6 &-0&.012758475(5)    &-0&.0078745(3)     &-0&.00280503(1)  &-0&.001180528(5)    &-0&.0008517(1)\\
7&$\times$1 &-0&.002884050(2)    &-0&.001612786(1)   &-0&.00050323(2)  &-0&.0002053982(6)   &-0&.00014676(1)\\
7&$\times$2 &-0&.006778774(7)    &-0&.003956439(1)   &-0&.00130823(4)  &-0&.000556193(2)    &-0&.00040694(4)\\
7&$\times$3 &-0&.00917979(1)     &-0&.005521988(2)   &-0&.00189010(4)  &-0&.000801642(3)    &-0&.00058404(8)\\
7&$\times$4 &-0&.01102290(2)     &-0&.006735038(6)   &-0&.00235759(3)  &-0&.000996348(4)    &-0&.00072237(9)\\
7&$\times$5 &-0&.01265212(2)     &-0&.00779486(1)    &-0&.002767215(5) &-0&.001165438(6)    &-0&.0008416(2)\\
7&$\times$6 &-0&.01418382(2)     &-0&.00878145(2)    &-0&.00314602(5)  &-0&.001321187(7)    &-0&.0009508(4)\\
7&$\times$7 &-0&.01566445(4)     &-0&.00972947(2)    &-0&.0035074(1)   &-0&.00146965(1)     &-0&.0010548(7)\\
8&$\times$1 &-0&.003292706(6)    &-0&.001841639(5)   &-0&.00057435(2)  &-0&.000233789(3)    &-0&.00016671(2)\\
8&$\times$2 &-0&.00768502(2)     &-0&.004489103(6)   &-0&.00148551(2)  &-0&.000630558(4)    &-0&.00046077(4)\\
8&$\times$3 &-0&.01031575(5)     &-0&.00621438(2)    &-0&.002131111(1) &-0&.000902522(5)    &-0&.00065662(4)\\
8&$\times$4 &-0&.01228369(8)     &-0&.00751941(3)    &-0&.00263960(4)  &-0&.00111385(1)     &-0&.0008065(1)\\
8&$\times$5 &-0&.0139944(1)      &-0&.00863994(3)    &-0&.0030781(1)   &-0&.00129432(3)     &-0&.0009331(1)\\
8&$\times$6 &-0&.0155865(1)      &-0&.00967139(3)    &-0&.0034790(2)   &-0&.00145857(2)     &-0&.0010476(2)\\
8&$\times$7 &-0&.0171157(2)      &-0&.01065545(6)    &-0&.0038584(3)   &-0&.00161390(3)     &-0&.0011559(2)\\
8&$\times$8 &-0&.0186086(3)      &-0&.0116122(2)     &-0&.0042254(6)   &-0&.00176401(6)     &-0&.0012604(3)\\

\end{tabular}
\end{minipage}
\end{center}
\end{table}

\begin{table}[ht]
%\linespread 1
\begin{center}
\begin{minipage}{17cm}
\caption{Values of \Red{{\bf $X^{b}_{SW}$}}, Eqs.(\ref{Wc}, \ref{Xi}), for specific masses.
\label{CloverSw2}}
\begin{tabular}{r@{}lr@{}lr@{}lr@{}lr@{}lr@{}l}
\multicolumn{2}{c}{$R\times T$}&
\multicolumn{2}{c}{$m=-0.8253968$} &
\multicolumn{2}{c}{$m=-0.5181059$} &
\multicolumn{2}{c}{$m=-0.1482168$} &
\multicolumn{2}{c}{$m=0.0$} &
\multicolumn{2}{c}{$m=0.038$} \\
\tableline \hline

 %M&$\times$N     m=-0.8253968          m=-0.5181059      m=-0.1482168            m=0.0               m=0.038
\\
1&$\times$1   &0&.000698193815(2)   &0&.00066565598(1)   &0&.0006181185(9)    &0&.00059630769(1)   &0&.0005903340(3)\\
2&$\times$1   &0&.00172584505(3)    &0&.00164432952(2)   &0&.001528663(3)     &0&.00147595111(2)   &0&.0014614713(8)\\
2&$\times$2   &0&.0038829733(2)     &0&.00370311435(3)   &0&.003455127(7)     &0&.00334255825(8)   &0&.003311403(2)\\
3&$\times$1   &0&.0026447690(1)     &0&.00251041847(3)   &0&.002327753(4)     &0&.00224630335(6)   &0&.0022240270(9)\\
3&$\times$2   &0&.0055686156(5)     &0&.0052917398(4)    &0&.00492817(2)      &0&.0047675835(2)    &0&.004723331(3)\\
3&$\times$3   &0&.007577888(1)      &0&.0071719960(7)    &0&.00666547(3)      &0&.0064495931(7)    &0&.006390427(3)\\
4&$\times$1   &0&.0035499739(3)     &0&.00336406864(7)   &0&.003114878(7)     &0&.0030047884(1)    &0&.002974759(4)\\
4&$\times$2   &0&.007181537(1)      &0&.0068154069(4)    &0&.00634099(3)      &0&.0061338378(5)    &0&.006076892(4)\\
4&$\times$3   &0&.009432576(3)      &0&.008914872(2)     &0&.00827670(4)      &0&.008008705(2)     &0&.007935550(4)\\
4&$\times$4   &0&.011436645(5)      &0&.010795193(5)     &0&.01001199(4)      &0&.009688495(3)     &0&.009600620(4)\\
5&$\times$1   &0&.0044532013(4)     &0&.00421590197(7)   &0&.00390024(2)      &0&.0037614497(2)    &0&.003723650(5)\\
5&$\times$2   &0&.008781297(2)      &0&.0083272219(9)    &0&.00774289(6)      &0&.007489215(1)     &0&.00741959(2)\\
5&$\times$3   &0&.011255252(4)      &0&.010628956(4)     &0&.00986167(6)      &0&.009541876(2)     &0&.00945480(2)\\
5&$\times$4   &0&.013384587(9)      &0&.012624976(8)     &0&.01170146(6)      &0&.011323119(5)     &0&.01122068(2)\\
5&$\times$5   &0&.01543030(1)       &0&.01454560(1)      &0&.01347306(7)      &0&.013037276(8)     &0&.01291972(5)\\
6&$\times$1   &0&.0053554122(5)     &0&.00506673408(7)   &0&.00468462(3)      &0&.0045171518(3)    &0&.004471594(5)\\
6&$\times$2   &0&.010375729(2)      &0&.0098339468(8)    &0&.00914016(4)      &0&.0088399114(7)    &0&.00875761(2)\\
6&$\times$3   &0&.013064833(5)      &0&.012330656(4)     &0&.01143529(5)      &0&.011063834(2)     &0&.01096286(2)\\
6&$\times$4   &0&.01530878(1)       &0&.014432416(8)     &0&.01337056(8)      &0&.012937683(5)     &0&.01282072(6)\\
6&$\times$5   &0&.01743948(1)       &0&.01643195(1)      &0&.01521349(9)      &0&.014720827(7)     &0&.01458826(9)\\
6&$\times$6   &0&.01951938(1)       &0&.01838393(1)      &0&.01701323(9)      &0&.016461641(8)     &0&.0163137(1)\\
7&$\times$1   &0&.006256861(1)      &0&.0059168274(1)    &0&.00546839(4)      &0&.0052721785(4)    &0&.005218838(8)\\
7&$\times$2   &0&.011966841(2)      &0&.0113374872(9)    &0&.01053447(5)      &0&.0101877335(4)    &0&.01009275(2)\\
7&$\times$3   &0&.014866656(3)      &0&.014024969(1)     &0&.01300215(6)      &0&.012579175(2)     &0&.01246432(5)\\
7&$\times$4   &0&.017219108(3)      &0&.016226710(2)     &0&.01502766(7)      &0&.014540534(4)     &0&.01440911(7)\\
7&$\times$5   &0&.019427331(3)      &0&.018298159(3)     &0&.01693557(7)      &0&.016386498(4)     &0&.0162390(2)\\
7&$\times$6   &0&.02156949(1)       &0&.02030783(1)      &0&.01878745(8)      &0&.018177603(4)     &0&.0180143(2)\\
7&$\times$7   &0&.02367264(5)       &0&.02228083(1)      &0&.0206060(1)       &0&.01993630(3)      &0&.0197574(2)\\
8&$\times$1   &0&.007157780(5)      &0&.006766415(2)     &0&.00625165(5)      &0&.006026753(2)     &0&.005965640(8)\\
8&$\times$2   &0&.013555778(5)      &0&.012838962(4)     &0&.01192690(5)      &0&.011533716(8)     &0&.01142607(1)\\
8&$\times$3   &0&.016663537(6)      &0&.015714608(9)     &0&.01456474(6)      &0&.014090370(8)     &0&.01396167(2)\\
8&$\times$4   &0&.019120705(7)      &0&.01801277(1)      &0&.01667727(9)      &0&.01613611(1)      &0&.0159903(8) \\
8&$\times$5   &0&.02140179(1)       &0&.02015178(5)      &0&.0186462(1)       &0&.01804107(2)      &0&.0178788(8)\\
8&$\times$6   &0&.02360086(4)       &0&.02221416(6)      &0&.0205457(1)       &0&.01987809(2)      &0&.0196997(9)\\
8&$\times$7   &0&.02575120(5)       &0&.02423078(6)      &0&.0224037(1)       &0&.02167474(2)      &0&.021480(1)\\
8&$\times$8   &0&.02787067(6)       &0&.02621858(7)      &0&.0242355(1)       &0&.0234460(1)       &0&.023235(1)\\

\end{tabular}
\end{minipage}
\end{center}
\end{table}

\begin{table}[ht]
%\linespread 1
\begin{center}
\begin{minipage}{14cm}
\caption{A comparison of our results for \Red{{\bf $X_{W}$}} with those of F. Di Renzo and L. Scorzato [10], and G. S. Bali and P. Boyle [11], for $m=0$.
\label{Comp1}}
\begin{tabular}{r@{}lr@{}lr@{}lr@{}l}
\multicolumn{2}{c}{$R\times T$}&
\multicolumn{2}{c}{This work} &
\multicolumn{2}{c}{Ref.[10]} &
\multicolumn{2}{c}{Ref.[11]} \\
\tableline \hline

  %M*N           Our results         DR&LS results    GB&PB results
\\
1&$\times$1 &0&.0006929202(2)   &0&.000688(3)     &0&.000696(2)\\
2&$\times$1 &0&.0013187555(4)   &0&.001310(8)     &0&.001326(3)\\
2&$\times$2 &0&.0023609937(7)   &0&.00235(2)      &0&.00238(1) \\
3&$\times$3 &0&.004385778(7)    &0&.00431(8)      &&        -  \\
4&$\times$4 &0&.006560910(5)    &0&.00615(19)     &&        -  \\
5&$\times$5 &0&.00882313(1)     &0&.00813(42)     &&        -  \\
6&$\times$6 &0&.01114489(4)     &0&.01010(71)     &&        -  \\
7&$\times$7 &0&.01351051(4)     &0&.0120(11)      &&        -  \\
8&$\times$8 &0&.0159099(2)      &0&.0133(18)      &&        -  \\

\end{tabular}
\end{minipage}
\end{center}
\end{table}

\begin{table}[ht]
%\linespread 1
\begin{center}
\begin{minipage}{9cm}
\caption{A comparison of our results for \Red{{\bf $X^a_{SW}$}} with those of G. S. Bali and P. Boyle [11], for $m=0$.
\label{Comp2}}
\begin{tabular}{r@{}lr@{}lr@{}l}
\multicolumn{2}{c}{$R\times T$}&
\multicolumn{2}{c}{This work} &
\multicolumn{2}{c}{Ref.[11]} \\
\tableline \hline
\\
1&$\times$1 &-0&.00002010061(1)  &-0&.0000202(3)\\
2&$\times$1 &-0&.0000570082(3)   &-0&.0000565(5)\\
2&$\times$2 &-0&.0001607273(9)   &-0&.000161(1)\\

\end{tabular}
\end{minipage}
\end{center}
\end{table}

\begin{table}[ht]
%\linespread 1
\begin{center}
\begin{minipage}{9cm}
\caption{A comparison of our results of \Red{{\bf $X^b_{SW}$}}  with those of G. S. Bali and P. Boyle [11], for $m=0$.
\label{Comp3}}
\begin{tabular}{r@{}lr@{}lr@{}l}
\multicolumn{2}{c}{$R\times T$}&
\multicolumn{2}{c}{This work} &
\multicolumn{2}{c}{Ref.[11]} \\
\tableline \hline
\\
1&$\times$1 &0&.00059630769(1)   &0&.00059635(1)\\
2&$\times$1 &0&.00147595111(2)   &0&.0014759(3) \\
2&$\times$2 &0&.00334255825(8)   &0&.0033420(5) \\

\end{tabular}
\end{minipage}
\end{center}
\end{table}

\begin{table}[ht]
%\linespread 1
\begin{center}
\begin{minipage}{17cm}
\caption{Values of \Mulberry{{\bf $X_{Overlap}$}}, Eq.(\ref{Xoverlap}), for specific $M_O$ values.
\label{Overlap}}
\begin{tabular}{r@{}lr@{}lr@{}lr@{}lr@{}lr@{}l}
\multicolumn{2}{c}{$R\times T$}&
\multicolumn{2}{c}{$M_O=0.2$} &
\multicolumn{2}{c}{$M_O=0.6$} &
\multicolumn{2}{c}{$M_O=1.0$} &
\multicolumn{2}{c}{$M_O=1.4$} &
\multicolumn{2}{c}{$M_O=1.8$} \\
\tableline \hline

\\
  %M*N           m=0.2              m=0.6              m=1.0               m=1.4              m=1.8
1&$\times$1 &-0&.00039465(1) &-0&.00056736986(6) &-0&.00070755250(4) &-0&.00079691506(6) &-0&.0008216755(4)\\
2&$\times$1 &-0&.00074225(3) &-0&.0010460401(2)  &-0&.0012643519(2)  &-0&.001379382(1)   &-0&.001388660(4)\\
2&$\times$2 &-0&.00130320(3) &-0&.0017788121(6)  &-0&.0020631095(7)  &-0&.002175237(1)   &-0&.00215719(3)\\
3&$\times$1 &-0&.00106622(3) &-0&.0014872672(5)  &-0&.0017747956(4)  &-0&.001914135(2)   &-0&.00191097(1)\\
3&$\times$2 &-0&.00178572(3) &-0&.0023943560(8)  &-0&.002728920(2)   &-0&.002844382(2)   &-0&.00281328(8)\\
3&$\times$3 &-0&.0023584(2)  &-0&.003088689(3)   &-0&.003451232(3)   &-0&.003559630(2)   &-0&.00352843(9)\\
4&$\times$1 &-0&.00138312(3) &-0&.0019189679(6)  &-0&.0022756915(8)  &-0&.002440160(3)   &-0&.00242470(3)\\
4&$\times$2 &-0&.0022438(1)  &-0&.002979068(2)   &-0&.003366361(3)   &-0&.003489302(6)   &-0&.00344688(7)\\
4&$\times$3 &-0&.0028839(5)  &-0&.003725538(3)   &-0&.004122705(5)   &-0&.004232257(7)   &-0&.0042034(2)\\
4&$\times$4 &-0&.0034499(5)  &-0&.004385664(7)   &-0&.004804389(5)   &-0&.00491307(2)    &-0&.0049016(3)\\
5&$\times$1 &-0&.00169775(3) &-0&.0023482303(6)  &-0&.002774573(1)   &-0&.002964467(3)   &-0&.00293665(5)\\
5&$\times$2 &-0&.0026941(4)  &-0&.003555500(3)   &-0&.003997457(4)   &-0&.004129009(3)   &-0&.0040746(2)\\
5&$\times$3 &-0&.0033921(5)  &-0&.004346425(4)   &-0&.004782379(4)   &-0&.00489521(2)    &-0&.0048676(2)\\
5&$\times$4 &-0&.0039898(6)  &-0&.005021256(4)   &-0&.005468337(6)   &-0&.00557929(3)    &-0&.0055829(2)\\
5&$\times$5 &-0&.0045506(7)  &-0&.00566255(1)    &-0&.006130452(9)   &-0&.00624369(6)    &-0&.0062747(2)\\
6&$\times$1 &-0&.00201153(7) &-0&.002776797(1)   &-0&.003272973(2)   &-0&.003488349(3)   &-0&.00344799(8)\\
6&$\times$2 &-0&.0031406(4)  &-0&.004129465(2)   &-0&.004626948(2)   &-0&.004767304(3)   &-0&.0047001(2)\\
6&$\times$3 &-0&.0038940(4)  &-0&.004962410(3)   &-0&.005438978(3)   &-0&.00555545(2)    &-0&.0055275(3)\\
6&$\times$4 &-0&.0045189(4)  &-0&.00564913(1)    &-0&.006127581(2)   &-0&.00624132(5)    &-0&.0062572(5)\\
6&$\times$5 &-0&.0050962(5)  &-0&.00629315(2)    &-0&.00678613(2)    &-0&.00690240(9)    &-0&.0069568(9)\\
6&$\times$6 &-0&.0056530(6)  &-0&.00692336(2)    &-0&.00743660(2)    &-0&.0075559(4)     &-0&.0076436(9)\\
7&$\times$1 &-0&.0023249(1)  &-0&.003205140(1)   &-0&.003771247(3)   &-0&.00401211(9)    &-0&.0039591(1)\\
7&$\times$2 &-0&.0035857(4)  &-0&.004702613(2)   &-0&.005256009(3)   &-0&.00540519(9)    &-0&.0053248(2)\\
7&$\times$3 &-0&.0043928(4)  &-0&.005576718(2)   &-0&.006094720(8)   &-0&.0062148(1)     &-0&.0061856(3)\\
7&$\times$4 &-0&.0050434(4)  &-0&.00627435(2)    &-0&.00678550(2)    &-0&.0069021(1)     &-0&.0069285(3)\\
7&$\times$5 &-0&.0056347(5)  &-0&.00691997(2)    &-0&.00744009(2)    &-0&.0075593(3)     &-0&.0076344(4)\\
7&$\times$6 &-0&.0062004(5)  &-0&.00754866(3)    &-0&.00808495(4)    &-0&.0082071(5)     &-0&.0083245(7)\\
7&$\times$7 &-0&.0067540(6)  &-0&.00817135(3)    &-0&.00872727(5)    &-0&.0088536(5)     &-0&.009007(1)\\
8&$\times$1 &-0&.0026381(1)  &-0&.003633404(3)   &-0&.004269491(3)   &-0&.00453583(9)    &-0&.0044702(1)\\
8&$\times$2 &-0&.0040302(4)  &-0&.005275469(5)   &-0&.005884963(7)   &-0&.00604296(2)    &-0&.0059491(2)\\
8&$\times$3 &-0&.0048902(4)  &-0&.006190440(5)   &-0&.00675029(2)    &-0&.00687404(6)    &-0&.0068426(2)\\
8&$\times$4 &-0&.0055652(5)  &-0&.006898582(6)   &-0&.00744318(2)    &-0&.0075626(1)     &-0&.0075982(2)\\
8&$\times$5 &-0&.0061694(6)  &-0&.007545429(7)   &-0&.00809377(5)    &-0&.0082150(2)     &-0&.0083097(6)\\
8&$\times$6 &-0&.0067427(6)  &-0&.008172214(9)   &-0&.00873296(6)    &-0&.0088588(5)     &-0&.009002(1)\\
8&$\times$7 &-0&.0073012(6)  &-0&.00879187(2)    &-0&.00936927(7)    &-0&.0094986(9)     &-0&.009684(2)\\
8&$\times$8 &-0&.0078520(8)  &-0&.00940894(1)    &-0&.01000531(9)    &-0&.010138(1)      &-0&.010363(2)\\

\end{tabular}
\end{minipage}
\end{center}
\end{table}

\begin{table}[ht]
%\linespread 1
\begin{center}
\begin{minipage}{15cm}
\caption{Total values of \Brown{{\bf $V_W$}}, \Brown{{\bf $V^{a}_{SW}$}} and \Brown{{\bf $V^{b}_{SW}$}}, Eq.(\ref{shiftclover}), for various bare fermion masses.
\label{InfClover}}
\begin{tabular}{r@{}lr@{}lr@{}lr@{}l}
\multicolumn{2}{c}{$m$}&
\multicolumn{2}{c}{$V_W$} &
\multicolumn{2}{c}{$V^{a}_{SW}$} &
\multicolumn{2}{c}{$V^{b}_{SW}$} \\
\tableline \hline
\\
  %Mass           $V_W$          $V^{a}_{SW}$        $V^{b}_{SW}$
-1&.014925   &0&.0034903(4)    &-0&.00196733(7)   &0&.00231391(3)\\
-0&.9512196  &0&.0033381(1)    &-0&.0018394(1)    &0&.00228645(1)\\
-0&.8749999  &0&.0031618(2)    &-0&.00168960(1)   &0&.00225268(4)\\
-0&.8253968  &0&.0030498(1)    &-0&.00159403(3)   &0&.00223031(4)\\
-0&.7948719  &0&.0029818(1)    &-0&.0015358(2)    &0&.00221643(3)\\
-0&.5181059  &0&.0023900(8)    &-0&.0010308(6)    &0&.0020891(1)\\
-0&.423462   &0&.0021944(9)    &-0&.0008671(8)    &0&.00204579(7)\\
-0&.4157708  &0&.002179(1)     &-0&.000854(1)     &0&.00204228(6)\\
-0&.4028777  &0&.0021523(4)    &-0&.000831(1)     &0&.00203642(2)\\
-0&.3140433  &0&.001974(3)     &-0&.000696(2)     &0&.0019965(4)\\
-0&.3099631  &0&.001966(2)     &-0&.000685(7)     &0&.0019947(5)\\
-0&.301775   &0&.001951(6)     &-0&.000663(4)     &0&.0019914(6)\\
-0&.2962964  &0&.001939(2)     &-0&.000653(2)     &0&.0019893(6)\\
-0&.2852897  &0&.001915(6)     &-0&.000634(2)     &0&.0019844(5)\\
-0&.2825278  &0&.001909(7)     &-0&.000629(1)     &0&.0019828(6)\\
-0&.2769916  &0&.001898(3)     &-0&.000620(1)     &0&.0019798(5)\\
-0&.2686568  &0&.001883(6)     &-0&.000607(4)     &0&.0019760(2)\\
-0&.1482168  &0&.001634(8)     &-0&.000401(4)     &0&.0019227(2)\\
0&.0         &0&.00134096(5)   &-0&.0001641(1)    &0&.00185871(2)\\
0&.005       &0&.0013266(1)    &-0&.0001561(2)    &0&.00185656(1)\\
0&.01        &0&.0013112(8)    &-0&.0001481(3)    &0&.001854398(7)\\
0&.014       &0&.0012982(9)    &-0&.0001420(1)    &0&.001852668(1)\\
0&.016       &0&.001291(1)     &-0&.0001389(1)    &0&.001851801(2)\\
0&.018       &0&.001284(2)     &-0&.0001360(2)    &0&.001850934(5)\\
0&.0236      &0&.0012623(4)    &-0&.0001279(2)    &0&.00184850(1)\\
0&.027       &0&.0012499(7)    &-0&.0001235(2)    &0&.00184701(1)\\
0&.035       &0&.001226(1)     &-0&.0001127(4)    &0&.00184349(4)\\
0&.0366      &0&.001218(4)     &-0&.0001105(5)    &0&.00184278(4)\\
0&.038       &0&.001212(4)     &-0&.0001087(6)    &0&.00184216(4)\\
0&.0427      &0&.001194(4)     &-0&.0001032(2)    &0&.00184007(4)\\
0&.046       &0&.001183(5)     &-0&.000100(1)     &0&.00183859(4)\\
0&.0535      &0&.001155(3)     &-0&.000092(2)     &0&.00183522(4)\\
0&.055       &0&.001150(3)     &-0&.000091(2)     &0&.00183454(4)\\
0&.072       &0&.001100(3)     &-0&.000073(3)     &0&.00182676(4)\\
0&.0927      &0&.001052(5)     &-0&.0000615(9)    &0&.00181709(5)\\

\end{tabular}
\end{minipage}
\end{center}
\end{table}

\begin{table}[ht]
%\linespread 1
\begin{center}
\begin{minipage}{8cm}
\caption{\Brown{{\bf $V_{Overlap}$}}, Eq.(\ref{shiftoverlap}), as a function of $M_O$.
\label{InfOverlap}}
\begin{tabular}{r@{}lr@{}l}
\multicolumn{2}{c}{$M_O$}&
\multicolumn{2}{c}{$V_{Overlap}$} \\
\tableline \hline
\\
0&.01 &-0&.0004974(4) \\
0&.05 &-0&.000477(4) \\
0&.1  &-0&.000483(2) \\
0&.2  &-0&.000517(3) \\
0&.3  &-0&.000526(2) \\
0&.4  &-0&.000533(1) \\
0&.5  &-0&.000546(1) \\
0&.6  &-0&.0005543(4) \\
0&.7  &-0&.0005617(1) \\
0&.8  &-0&.00056556(7) \\
0&.9  &-0&.00056756(2) \\
1&.   &-0&.00056833(7) \\
1&.1  &-0&.00056827(3) \\
1&.2  &-0&.0005681(4) \\
1&.3  &-0&.0005690(1) \\
1&.4  &-0&.000571(1) \\
1&.5  &-0&.000573(3) \\
1&.6  &-0&.000591(3) \\
1&.7  &-0&.000612(5) \\
1&.8  &-0&.000651(3) \\
1&.9  &-0&.000629(3) \\
1&.95 &-0&.000649(3) \\
1&.99 &-0&.000616(1) \\

\end{tabular}
\end{minipage}
\end{center}
\end{table}

\begin{table}[ht]
%\linespread 1
\begin{center}
\begin{minipage}{14cm}
\caption{\PineGreen{{\bf $F_{Clover}\left(R\right)$}}, Eq.(\ref{pot}), for specific $m$ values (\Blue{$c_{SW}=1.3$}).
\label{CloverPotential}}
\begin{tabular}{r@{}lr@{}lr@{}lr@{}l}
\multicolumn{2}{c}{$R$}&
\multicolumn{2}{c}{$m=-0.5181059$} &
\multicolumn{2}{c}{$m=0.0$} &
\multicolumn{2}{c}{$m=0.038$} \\
\tableline \hline
\\
&1  &0&.00232276(8) &0&.00180277(2)&0&.0017638(1)\\
&2  &0&.0035644(3)  &0&.0030071(2) &0&.0029552(2)\\
&3  &0&.0039016(5)  &0&.0033731(2) &0&.0033175(2)\\
&4  &0&.0040550(7)  &0&.0035566(3) &0&.0034997(7)\\
&5  &0&.0041511(9)  &0&.0036759(4) &0&.003618(1)\\
&6  &0&.0042185(9)  &0&.0037608(5) &0&.003702(1)\\
&7  &0&.004268(1)   &0&.0038242(9) &0&.003765(3)\\
&8  &0&.00431(1)    &0&.003873(6)  &0&.003814(4)\\
&9  &0&.00434(3)    &0&.00393(2)   &0&.00387(2)\\
&10 &0&.00436(3)    &0&.00395(2)   &0&.00389(2)\\
&11 &0&.00437(5)    &0&.00397(4)   &0&.00391(3)\\
&12 &0&.00439(11)   &0&.00398(6)   &0&.00393(5)\\

\end{tabular}
\end{minipage}
\end{center}
\end{table}

\begin{table}[ht]
\begin{center}
\begin{minipage}{17cm}
\caption{{\bf $F_{Overlap}\left(R\right)$} for specific $M_O$ values.
\label{OverlapPotential}}
\begin{tabular}{r@{}lr@{}lr@{}lr@{}lr@{}l}
\multicolumn{2}{c}{$R$}&
\multicolumn{2}{c}{$M_O=0.5$} &
\multicolumn{2}{c}{$M_O=1.0$} &
\multicolumn{2}{c}{$M_O=1.3$} &
\multicolumn{2}{c}{$M_O=1.5$} \\
\tableline \hline
\\
&1   &-0&.00040364(3)   &-0&.00049870(6)  &-0&.00052178(1) &-0&.00052476(4)\\
&2   &-0&.0005487(2)    &-0&.00063045(9)  &-0&.00064031(9) &-0&.00063809(4)\\
&3   &-0&.0005938(2)    &-0&.0006583(2)   &-0&.0006635(2)  &-0&.00066247(7)\\
&4   &-0&.0006078(2)    &-0&.0006619(2)   &-0&.0006659(5)  &-0&.0006674(5)\\
&5   &-0&.0006114(7)    &-0&.0006598(5)   &-0&.000663(1)   &-0&.0006664(6)\\
&6   &-0&.0006121(7)    &-0&.0006563(6)   &-0&.000659(2)   &-0&.0006644(7)\\
&7   &-0&.000611(1)     &-0&.000650(1)    &-0&.000656(2)   &-0&.0006619(7)\\
&8   &-0&.000609(1)     &-0&.000648(2)    &-0&.000653(2)   &-0&.000659(1)\\
&9   &-0&.000607(1)     &-0&.000646(2)    &-0&.000651(2)   &-0&.000657(2)\\
&10  &-0&.000607(2)     &-0&.000644(3)    &-0&.000648(3)   &-0&.000655(5)\\
&11  &-0&.000606(3)     &-0&.000640(3)    &-0&.000646(4)   &-0&.000656(6)\\
&12  &-0&.000608(3)     &-0&.000639(6)    &-0&.000642(4)   &-0&.000653(7)\\

\end{tabular}
\end{minipage}
\end{center}
\end{table}

\begin{table}[ht]
\begin{center}
\begin{minipage}{17cm}
\caption{{\bf $F_{W}\left(R\right)$}, for specific values of the bare fermion mass $m$.
\label{CloverVSw0}}
\begin{tabular}{r@{}lr@{}lr@{}lr@{}lr@{}lr@{}l}
\multicolumn{2}{c}{$R$}&
\multicolumn{2}{c}{$m=-0.8253968$} &
\multicolumn{2}{c}{$m=-0.5181059$} &
\multicolumn{2}{c}{$m=-0.1482168$} &
\multicolumn{2}{c}{$m=0.0$} &
\multicolumn{2}{c}{$m=0.038$} \\
\tableline \hline

\\
&1    &0&.001610803(2)  &0&.00118506(4)   &0&.0007257(9)   &0&.00056593(1)   &0&.00052922(6)\\
&2    &0&.0022839(2)    &0&.00172190(7)   &0&.001077(2)    &0&.0008350(2)    &0&.0007782(1)\\
&3    &0&.0025550(2)    &0&.0019519(2)    &0&.001243(2)    &0&.0009637(2)    &0&.0008954(3)\\
&4    &0&.0026858(3)    &0&.0020665(2)    &0&.001335(2)    &0&.0010362(3)    &0&.0009615(3)\\
&5    &0&.0027612(6)    &0&.0021333(2)    &0&.001396(3)    &0&.0010831(3)    &0&.0010040(7)\\
&6    &0&.002810(1)     &0&.0021769(4)    &0&.001440(5)    &0&.0011163(3)    &0&.001034(1)\\
&7    &0&.002845(3)     &0&.002208(8)     &0&.001465(5)    &0&.0011412(5)    &0&.001056(2)\\
&8    &0&.00288(1)      &0&.002235(8)     &0&.001485(8)    &0&.001161(2)     &0&.001073(6)\\
&9    &0&.00289(5)      &0&.002249(8)     &0&.001500(9)    &0&.001179(8)     &0&.001089(8)\\
&10   &0&.00291(6)      &0&.002263(8)     &0&.00153(3)     &0&.001192(9)     &0&.001102(9)\\
&11   &0&.00290(7)      &0&.00227(1)      &0&.00152(7)     &0&.001202(9)     &0&.001108(9)\\
&12   &0&.00292(7)      &0&.00228(4)      &0&.00153(7)     &0&.00121(2)      &0&.00112(1)\\

\end{tabular}
\end{minipage}
\end{center}
\end{table}

\begin{table}[ht]
\begin{center}
\begin{minipage}{17cm}
\caption{{\bf $F^a_{SW}\left(R\right)$}, for specific values of the bare fermion mass $m$. 
\label{CloverVSw1}}
\begin{tabular}{r@{}lr@{}lr@{}lr@{}lr@{}lr@{}l}
\multicolumn{2}{c}{$R$}&
\multicolumn{2}{c}{$m=-0.8253968$} &
\multicolumn{2}{c}{$m=-0.5181059$} &
\multicolumn{2}{c}{$m=-0.1482168$} &
\multicolumn{2}{c}{$m=0.0$} &
\multicolumn{2}{c}{$m=0.038$} \\
\tableline \hline

\\
&1 &-0&.00040723(2)  &-0&.00022779(3)   &-0&.0000709(6)   &-0&.000028181(3)  &-0&.00001984(5)\\
&2 &-0&.00090060(7)  &-0&.0005284(1)    &-0&.000176(2)    &-0&.00007354(1)   &-0&.0000533(1)\\
&3 &-0&.00112346(9)  &-0&.0006831(2)    &-0&.000238(3)    &-0&.00009913(2)   &-0&.0000714(1)\\
&4 &-0&.0012390(1)   &-0&.0007681(2)    &-0&.000276(4)    &-0&.00011436(3)   &-0&.0000821(2)\\
&5 &-0&.0013092(1)   &-0&.0008204(4)    &-0&.000301(6)    &-0&.00012411(4)   &-0&.0000886(3)\\
&6 &-0&.0013564(1)   &-0&.0008555(4)    &-0&.000324(8)    &-0&.00013073(7)   &-0&.0000931(3)\\
&7 &-0&.001390(1)    &-0&.0008805(5)    &-0&.000339(9)    &-0&.00013549(7)   &-0&.0000957(5)\\
&8 &-0&.001416(4)    &-0&.000899(2)     &-0&.000340(9)    &-0&.0001395(9)    &-0&.0000976(5)\\
&9 &-0&.00144(1)     &-0&.000916(7)     &-0&.000344(9)    &-0&.0001419(9)    &-0&.000099(1)\\
&10&-0&.00145(1)     &-0&.000927(8)     &-0&.000351(9)    &-0&.000144(1)     &-0&.000100(1)\\
&11&-0&.00146(1)     &-0&.000940(8)     &-0&.000356(9)    &-0&.000146(2)     &-0&.000102(2)\\
&12&-0&.00147(4)     &-0&.00094(3)      &-0&.000357(9)    &-0&.000147(6)     &-0&.000102(3)\\

\end{tabular}
\end{minipage}
\end{center}
\end{table}

\begin{table}[ht]
\begin{center}
\begin{minipage}{17cm}
\caption{{\bf $F^b_{SW}\left(R\right)$}, for specific values of the bare fermion mass $m$.
\label{CloverVSw2}}
\begin{tabular}{r@{}lr@{}lr@{}lr@{}lr@{}lr@{}l}
\multicolumn{2}{c}{$R$}&
\multicolumn{2}{c}{$m=-0.8253968$} &
\multicolumn{2}{c}{$m=-0.5181059$} &
\multicolumn{2}{c}{$m=-0.1482168$} &
\multicolumn{2}{c}{$m=0.0$} &
\multicolumn{2}{c}{$m=0.038$} \\
\tableline \hline
\\

&1   &0&.000899650(2)  &0&.000848415(3) &0&.00078216(4) &0&.0007535343(8) &0&.000745769(9)\\
&2   &0&.001583856(4)  &0&.001496742(3) &0&.0013881(1)  &0&.001341834(6)  &0&.00132920(1)\\
&3   &0&.00178552(2)   &0&.001679062(4) &0&.0015522(5)  &0&.00150193(2)   &0&.00148814(4)\\
&4   &0&.00188161(6)   &0&.001767482(4) &0&.0016328(5)  &0&.00157932(2)   &0&.00156503(8)\\
&5   &0&.00194374(6)   &0&.00182508(6)  &0&.0016848(8)  &0&.00162962(2)   &0&.0016150(1)\\
&6   &0&.00198803(6)   &0&.00186610(2)  &0&.0017218(8)  &0&.00166535(5)   &0&.0016504(2)\\
&7   &0&.0020210(2)    &0&.0018965(2)   &0&.0017493(8)  &0&.0016918(2)    &0&.0016767(3)\\
&8   &0&.0020462(8)    &0&.001920(1)    &0&.001770(2)   &0&.001712(2)     &0&.001697(1)\\
&9   &0&.00207(1)      &0&.00194(1)     &0&.00180(1)    &0&.00173(1)      &0&.00172(1)\\
&10  &0&.00208(1)      &0&.00195(1)     &0&.00180(1)    &0&.00174(1)      &0&.00173(1)\\
&11  &0&.00210(2)      &0&.00197(1)     &0&.00181(3)    &0&.00175(2)      &0&.00174(2)\\
&12  &0&.00210(2)      &0&.00197(2)     &0&.00182(3)    &0&.00176(2)      &0&.00174(2)\\

\end{tabular}
\end{minipage}
\end{center}
\end{table}


\begin{references}

\bibitem{AXEGHAK} A. X. El-Khadra, G. Hockney, A. S. Kronfeld 
and P. B. Mackenzie, Phys. Rev. Lett.{\bf 69}, 729 (1992).

\bibitem{CTDKH1} C. T. Davies, K. Hornbostel, G. P. Lepage, A. Lidsey, 
J. Shigemitsu and J. H. Sloan, Phys. Lett. B {\bf 345}, 42 (1995)
[arXiv:hep-ph/9408328].

\bibitem{CTDKH2} C. T. Davies, K. Hornbostel, G. P. Lepage, P. McCallum, 
J. Shigemitsu and J. H. Sloan, Phys. Rev. D {\bf 56}, 2755 (1997)
[arXiv:hep-lat/9703010].

\bibitem{ASEA} A. Spitz {\it et al.}  [T$\chi$L Collaboration],
Phys. Rev. D {\bf 60}, 074502 (1999)
[arXiv:hep-lat/9906009]. 

\bibitem{SBEA} S. Booth {\it et al.}  [QCDSF and UKQCD collaborations],
Phys.\ Lett.\ B {\bf 519}, 229 (2001)
[arXiv:hep-lat/0103023]; Nucl.\ Phys.\ Proc.\ Suppl.\  {\bf 106}, 308 (2002)
[arXiv:hep-lat/0111006].

\bibitem{CDEA} C. Davies {\it et al.}, Nucl. Phys. B (PS) {\bf 119}, 595 (2003) 
[arXiv:hep-lat/0209122].

\bibitem{GP} G. Parisi,
in ``Proc. of the 20th Int. Conf. on High Energy Physics'',
Madison, Jul 17-23, 1980, eds. L. Durand
and L. G. Pondrom, (American Inst. of Physics, New York, 1981).

\bibitem{GPLPBM} G. P. Lepage and P. B. Mackenzie,
Phys. Rev. D {\bf 48}, 2250 (1993)
[arXiv:hep-lat/9209022].

\bibitem{GMCTS} G. Martinelli and C. T. Sachrajda, 
Nucl.Phys. {\bf B559}, 429 (1999) [arXiv:hep-lat/9812001]. 

\bibitem{DRLS} F. Di Renzo and L. Scorzato, JHEP {\bf 0410}, 073 (2004) [arXiv:hep-lat/0410010].

\bibitem{GBPB} G. Bali and P. Boyle, hep-lat/0210033.

\end{references}
\end{document}